\documentclass[12pt]{article}
\usepackage{amsfonts}

\renewcommand{\theequation}{\arabic{section}.\arabic{equation}}
\begin{document}
\thispagestyle{empty}

\begin{flushright}
  November 8, 2010 \\
 V2. December 15, 2010 \\
 V3. March 22, 2011
\end{flushright}

\vskip 2.5cm

\begin{center}
\baselineskip=16pt
 {\Large\bf  Superfield equations for the interacting system of}\\  \vskip 0.3cm
{\Large\bf $D=4$, ${\cal N}=1$  supermembrane and scalar multiplet }

\vskip 1.5cm

 {\large\bf
 Igor A. Bandos $^\dagger$$^\ast$ and
 Carlos Meliveo
 $^\ast$
 }
\vskip 1.0cm {\it\small $^\dagger$ IKERBASQUE, the Basque Foundation for Science, Bilbao,
Spain  \\  $^\ast$ Department of Theoretical Physics,
University of the Basque Country,  \\ P.O. Box 644, 48080 Bilbao,
Spain
}
  \vspace{24pt}

\end{center}

\vskip 0.5cm

\par
\begin{quote}

We present the superfield action for the  dynamical ${\cal N}=1$ $D=4$ supermembrane
in interaction with a dynamical scalar multiplet and use it to derive the superfield equations of motion. These include the supermembrane equations, which formally coincide with equations of supermembrane in a background of the (off--shell) scalar multiplet, and the special chiral superfield equations with supermembrane source.
In the case when the scalar supermultiplet part of the action contains only the simplest kinetic term
we have also extracted the spacetime component field equations from the superfield equations
and solve these in the leading order on supermembrane tension.
The inclusion of nontrivial superpotential and relation with known supersymmetric domain wall solutions is briefly discussed.

\end{quote}

\newpage

\setcounter{page}{2}
\section{Introduction}
\setcounter{equation}{0}

The development of String and M-theory \cite{M-theory} demonstrated the important r\^{o}le of supersymmetric extended objects, including eleven dimensional supermembrane \cite{BST87}, called now M2-brane. The consistency of this model in curved 11D superspace requires the torsion and curvature of that to obey the supergravity constraints. These, in its turn, result in the supergravity equations of motion so that the supergravity dynamics is in a way governed by the M2-brane.

Soon after the pioneer papers \cite{BST87} the simplest nontrivial counterpart of M2-brane, the $D=4$, ${\cal N}=1$ supermembrane, was studied in \cite{4Dsuperm}. Its consistency in curved superspace also requires a set of superspace  constraints. However, in distinction to the 11D case, these $D=4$, ${\cal N}=1$ constraints are {\it off--shell} in the sense that they do not produce equations of motion as their consequences. This implies that it is possible to construct the manifestly supersymmetric superfield  Lagrangian description of the interacting system of $D=4$, ${\cal N}=1$ supergravity and supermembrane. Curiously enough, such a system has not been constructed yet. We intend to turn to this problem in the future paper. Here, as a first stage in this direction, we will construct the superfield Lagrangian description and obtain the superfield equations of motion for a simpler  interacting system of supermembrane and scalar supermultiplet.

The superfield Lagrangian approach to  $D=4$, ${\cal N}=1$ supergravity---superparticle dynamical system  was developed in \cite{BAIL2}. The superfield equations for the dynamical system of supergravity, superstring and tensorial supermultiplet were obtained in \cite{B+I03}.
Both models were used to study the origin and properties of the complete but gauge fixed Lagrangian description of the supergravity--super--$p$--brane system by the sum of the spacetime, component action for supergravity (without auxiliary fields) and the purely bosonic limit of the super--$p$--brane action. These description proposed and developed in \cite{BdAI1,BAIL2,B+I03,IB+JdA05} can be used also for higher dimensional supergravity plus brane(s) interacting systems. However, the study of lower dimensional superfield equations for the interacting system of supergravity and superbrane, when it is possible, is also of interest as it might provide new insights in the properties of more complicated M-theoretical systems, and also on its own, as a sector of possible phenomenologically interesting models constructed on the basis of  four dimensional supergravity.

The system of superfield equation for the interacting system of supergravity, superstring and tensor supermultiplet, which was obtained in \cite{B+I03}, happened to be too complicated to be practical. Probably, to obtain a less complex set of superfield equations one have to use the superfield formulation of the so--called new minimal supergravity \cite{new-mSG75-78,new-mSG81} instead of the 'old' minimal formulation
\cite{minSG,WessZumino77} used in \cite{B+I03}. This hope is related to the fact that the new minimal formulation includes an auxiliary antisymmetric tensor which has natural coupling to string model so that one would not need to introduce, as in \cite{B+I03}, tensorial multiplet in addition to supergravity one.

On the other hand, one can use the results of \cite{B+I03} to extract the superfield equations for the superstring interacting with tensorial multiplet in flat $D=4$, ${\cal N}=1$ superspace. The existence of such a nontrivial interaction is related to the fact that, according to \cite{Gates:1980ay}, tensorial multiplet can be used to construct a supersymmetric closed 3-form in the flat $D=4$, ${\cal N}=1$ superspace. It is natural to begin with the study of these tensorial multiplet--superstring interacting equations before passing to the superstring interacting with supergravity.

The system of interacting superfield equations  could be expected to  be even simpler if we were writing it for a supersymmetric extended object interacting with a scalar supermultiplet. This is not possible for superstring but is possible for supermembrane, because the three form field strength cannot be constructed from the scalar multiplet, but the four form field strength can. Thus, as we have already mentioned, the aim of our present study is to construct the Lagrangian description and to obtain superfield equations of motion for this $D=4$, ${\cal N}=1$ interacting system of supermembrane and scalar multiplet.

This paper is organized as follows. In section 2 we introduce our basic superspace notation and review the generic scalar multiplet action, superfield equations of motion and dual description by a three form potential. To be more precise about the last item, in sec. 2.3 we present the closed four form in $D=4$, ${\cal N}=1$ superspace which is constructed from generic scalar supermultiplet. The explicit form of the three form potential is given  later, in Sec. 4.1, and only for the case of special scalar supermultiplet. In Sec. 3 we present the $D=4$, ${\cal N}=1$ supermembrane action and obtain the supermembrane equations of motion in  the background of (an off-shell) scalar multiplet. In Sec. 4 we derive superfield equations of motion for the special scalar supermultiplet interacting with supermembrane. The supermembrane current superfield determining the
{\it r.h.s.}'s of these equations is presented in Sec. 4.3. Sec. 5 is devoted to extracting the spacetime component field equations from the superfield equations with supermembrane current. For the sake of simplicity the discussion in Sec.5 is restricted to the case where the field theoretical part of the action is given by a free kinetic term only. In sec. 5.3 we present a solution of these dynamical field equations at the leading order in supermembrane tension. In Sec. 6 we present our conclusions and also compare the coupling of the membrane to background  scalar field which follows from our superfield supermembrane action  with one studied in \cite{Tomas+}.
The section 6.1. ('Discussion added') briefly discuss the contribution from nontrivial superpotential and relation with the known domain wall solutions.
Appendix A collects some useful technical details. The complete expressions for the components of the supermembrane current superfields are collected in Appendix B.

\section{Scalar multiplet description by chiral superfield.  }
\setcounter{equation}{0}

\subsection{Basic notation.}

We denote the supervielbein of  ${\cal N}=1$, D=4  superspace $\Sigma^{(4|4)}$ by
\begin{eqnarray}\label{EA=EaEf}
{E}^{ {A}}:=
dz^{ {M}}E_{ {M}}{}^{ {A}}(z)=
 ({E}^{ {a}}, {E}^{ {\alpha}}, \bar{E}^{\dot{\alpha}})\; , \qquad \cases{{a}=0,1,2,3 \; , \cr  \alpha=1,2 \; , \cr  \dot{\alpha}=1,2 \; .   } \;  \qquad
\end{eqnarray}
When superspace is flat, these obeys the constraints
\begin{eqnarray}
\label{Ta=4D} & dE^{ {a}} =
-2i{E}\wedge \sigma^{ {a}} {\bar E}  \; , \qquad  dE^{ \alpha} = 0  \; ,  \qquad  d\bar{E}{}^{ \dot{\alpha}} = 0 \; ,
\end{eqnarray}
where $\sigma^a_{\alpha\dot{\alpha}}$ are relativistic Pauli matrices, $d$ is exterior derivative of differential forms ($dd=0$, $dE^A= d(dZ^ME_M^A(Z))= dZ^M\wedge dE_M^A(Z)= dZ^M\wedge dZ^N \partial_N E_M^A(Z) =dZ^M\wedge dZ^N {\partial E_M^A(Z)\over \partial Z^N}$ {\it etc.}) and $\wedge$ denotes the exterior product of the differential form. This is antisymmetric for bosonic one forms, $E^a\wedge E^b=- E^b\wedge E^a$, symmetric for two fermionic one forms, $E^{ \alpha} \wedge E^{ \beta} =E^{ \beta} \wedge E^{ \alpha}$, and again antisymmetric for the product of bosonic and fermionic one  forms,  $E^a\wedge E^{\beta}=- E^{ \beta}\wedge E^a$. The constraints (\ref{Ta=4D})  can be solved by
\begin{eqnarray}\label{Eadx-i}
{E}^{ {a}}= dx^a - i d\theta^{\alpha} \sigma^a_{\alpha\dot{\alpha}} \bar{\theta}{}^{\dot{\alpha}} + i\theta^{\alpha} \sigma^a_{\alpha\dot{\alpha}} d\bar{\theta}{}^{\dot{\alpha}}\; ,   \qquad E^{\alpha}=  d\theta^{\alpha}\; , \qquad \bar{E}{}^{\dot{\alpha}}=d\bar{\theta}{}^{\dot{\alpha}}\;\qquad
\end{eqnarray}
expressing the supervielbein in terms of superspace coordinates
\begin{eqnarray}\label{Z=x,th}
z^{{ {M}}}= (x^{a}\, ,
\theta^{{\alpha}}, \bar{\theta}{}^{\dot{{\alpha}}})\; , \quad
{{\alpha}}=1,2\; , \qquad \dot{{\alpha}}=1,2\; , \qquad { {a}}=0,1,2,3 .  \qquad
\end{eqnarray}
 Their set includes the commuting bosonic coordinates  $x^{a}$ ($x^ax^b=x^bx^a$) and anticommuting fermionic coordinates $\theta^{{\alpha}}$ and $\bar{\theta}{}^{\dot{{\alpha}}}=(\theta^{{\alpha}})^*$
($\theta^{{\alpha}}\theta^{{\beta}}=-\theta^{{\beta}} \theta^{{\alpha}}\;$, $\;\theta^{{\alpha}}x^a= x^a \theta^{{\alpha}}$ {\it etc.}).

Decomposing the exterior derivative on the supervielbein basis
\begin{eqnarray}\label{d=ED}
d= E^\alpha D_\alpha + \bar{E}{}^{\dot{\alpha}}\bar{D}_{\dot{\alpha}} + E^a D_a\; , \qquad
\end{eqnarray}
we obtain the expressions for supersymmetric covariant derivatives,
\begin{eqnarray} \label{Dalpha=}
 D_a=\partial_a \; , \qquad
 D_\alpha= \partial_\alpha + i (\sigma^a\bar{\theta})_\alpha \partial_a \; , \qquad
 \bar{D}_{\dot{\alpha}} = \bar{\partial}_{\dot{\alpha}} + i (\theta\sigma^a)_{\dot{\alpha}} \partial_a =- (D_\alpha)^* \; . \qquad
\end{eqnarray}
These obey the superalgebra with the only one nontrivial (anti-)commutation relation
\begin{eqnarray}\label{(DfbDf)=}
{}\{ D_\alpha ,  \bar{D}_{\dot{\alpha}} \}= 2i \sigma^a_{\alpha\dot{\alpha}} \partial_a \; . \qquad
\end{eqnarray}

The possibility to construct the $D=4$, ${\cal N}=1$ supermembrane action is related to that in $D=4$, ${\cal N}=1$ superspace these exists the following supersymmetric invariant closed 4-form\footnote{Notice that in our notation the exterior derivative acts from the right, so that for any $p$-form $\Omega_p$ and $q$-form $\Omega_q$, $\; d(\Omega_p\wedge \Omega_q)= \Omega_p\wedge d\Omega_q + (-)^q d\Omega_p\wedge \Omega_q$.}
\begin{eqnarray}\label{h4:=C-E}
h_4=dc_3 :=- {i\over 4} E^b\wedge E^a\wedge E^\alpha\wedge E^\beta \sigma_{ab}{}_{\alpha\beta} + {i\over 4} E^b\wedge E^a \wedge \bar{E}^{\dot{\alpha}}\wedge \bar{E}^{\dot{\beta}} \tilde{\sigma}_{ab}{}_{\dot{\alpha}\dot{\beta}}\; .  \qquad
\end{eqnarray}
This describes a 3-cocycle which is nontrivial in Chevalley-Eilenberg (CE) cohomology \cite{JdA+PKT89,Jose+Paul=PRL89}, which implies that $h_4$ is a supersymmetric invariant closed four form, $dh_4=0$ and, despite it can be expressed as an exterior derivative of a 3-form, $h_4=dc_3$ (and, hence, is trivial cocycle of de Rahm cohomology), the corresponding 3-form $c_3$ is not invariant under supersymmetry.

\subsection{Scalar supermultiplet as described by chiral superfield}\label{scalarM}
 \label{scalarSec}

In this section we review the well known description of scalar supermultiplet by chiral superfield in superspace \cite{Ogievetsky:1975nu,Bagger+Wess=92,1001}.

The simplest irreducible representation  of the  $D=4$, ${\cal N}=1$  supersymmetry,
the scalar supermultiplet, is described by the chiral superfield, this is to say by complex superfield obeying the so-called chirality equation
\begin{eqnarray}\label{bDP=0}
\bar{D}_{\dot{\alpha}}\Phi =0 \; . \qquad
\end{eqnarray}
The complex conjugate  ({\it c.c.}), $\overline{\Phi}= (\Phi)^*$, obeys
\begin{eqnarray}\label{DbP=0}
D_{{\alpha}}\overline{\Phi} =0 \;  \qquad
\end{eqnarray}
and is called anti--chiral superfield.  The free equations of motion for the physical fields of a massless scalar supermultiplet ($\phi(x)= \Phi\vert_{\theta=0}$ and  $i\psi_\alpha (x)= D_\alpha\Phi\vert_{\theta=0}$) are collected in the superfield equation
\begin{eqnarray}\label{DDP=0}
DD{\Phi}:= D^{{\alpha}}D_{{\alpha}}{\Phi}=0  \; . \qquad
\end{eqnarray}
These equations can be derived from the action
\begin{eqnarray}\label{S=PhibPhi}
S_{kin}= \int d^8z\, \Phi \bar{\Phi}  \; , \qquad
\end{eqnarray}
where the superspace integration measure $d^8z=d^4xd^2\theta d^2\bar{\theta}$ is normalized as\footnote{Notice that, although the {\it r.h.s.} of this equation  is not manifestly hermitian, its imaginary part is integral of complete derivative
(as far as $\bar{D}\bar{D}\, DD=  DD\bar{D}\bar{D}\, -4i \tilde{\sigma}{}^{a\, \dot{\alpha}\alpha } \partial_a[D_\alpha , \bar{D}_{\dot{\alpha}}]$)
and, as such, can be ignored in our discussion. }
\begin{eqnarray}\label{S=PhibPhi}
d^8z=d^4x \bar{D}\bar{D}\; DD\, := d^4x \bar{D}_{\dot{\alpha}}\bar{D}{}^{\dot{\alpha}}\; D^\alpha D_\alpha \, \; .
\end{eqnarray}
Indeed, the variation of this functional reads $\delta S_{kin}= \int d^8z\, (\Phi \delta \bar{\Phi} + \delta\Phi \,  \bar{\Phi} )$. As far as the variation of chiral superfield should be chiral, $\bar{D}_{\dot{\alpha}}\delta\Phi=0$, and ${D}_{{\alpha}}\delta\bar{\Phi}=0$, we can equivalently write the action variation as
$\delta S_{kin}= \int d^4x\, \bar{D}_{\dot{\alpha}} \bar{D}{}^{\dot{\alpha}} (( DD\Phi)  \delta \bar{\Phi}) + c.c.$, which results in the equations of motion (\ref{DDP=0}).

The most general selfinteraction of the scalar supermultiplet is described by the superfield action
\begin{eqnarray}\label{S=PhibPhi}
S_{s-int}{[\Phi ; \bar{\Phi}]}&=& \int d^8z\, {\cal K}(\Phi , \bar{\Phi}) +  \int d^6\zeta_L \, W(\Phi )  +
\int d^6\zeta_R \, \bar{W}(\bar{\Phi}) = \nonumber \\
&=& \int d^4x \, \bar{D}\bar{D}\, DD\; {\cal K}(\Phi , \bar{\Phi}) +  \int d^4x DD \, W(\Phi )  + \int d^4x \, \bar{D}\bar{D}\,  \, \bar{W}(\bar{\Phi})
 \;  , \qquad
\end{eqnarray}
where   ${\cal K}(\Phi , \bar{\Phi})$ is an  arbitrary function of chiral superfield and its complex conjugate called {\it K\"ahler potential} and $W(\Phi)$ ($=(\bar{W}(\bar{\Phi})^*$) is an arbitrary holomorphic function of the complex scalar superfield $\Phi$ called {\it superpotential}. This latter is chiral, $\bar{D}_{\dot\alpha}W(\Phi)=W^\prime(\Phi)\bar{D}_{\dot\alpha}\Phi =0$, and hence is integrated with chiral measure defined by $d^6\zeta_L =d^4x DD$ (and $d^6\zeta_R =d^4x \bar{D}\bar{D}$).
To have the standard kinetic term for the scalar field of the supermultiplet, the K\"ahler potential is usually chosen to obey
 \begin{eqnarray}\label{K''=}
{\cal K}^{\prime\prime}_{\varphi \,  \bar{\varphi}}(\varphi , \bar{\varphi}):=  {\partial \over \partial \varphi}  {\partial \over \partial \bar{\varphi}}{\cal K}(\varphi , \bar{\varphi})\not=0\; .  \qquad
\end{eqnarray}

The superfield equations of motion following from the action $S_{s-int}[\Phi ; \bar{\Phi}]$ (\ref{S=PhibPhi}) are
\begin{eqnarray}\label{EqM=sc0b}
\bar{{\cal E}}&:= & DD {\cal K}^\prime_{\bar{\Phi}} +  \bar{W}^\prime_{\bar{\Phi} } = \qquad \nonumber \\  &=& DD\Phi \; {\cal K}^{\prime\prime}_{\Phi \bar{\Phi} }(\Phi , \bar{\Phi}) +
D^\alpha\Phi \, D_\alpha\Phi \; {\cal K}^{\prime\prime\prime}_{\Phi  \Phi \bar{\Phi}}(\Phi , \bar{\Phi}) + \bar{W}^\prime_{\bar{\Phi} } (\bar{\Phi})  =0 \; ,
 \qquad \\ \label{EqM=sc0}
{{\cal E}}&:= & \bar{D}\bar{D} {\cal K}^{\prime}_{{\Phi}} +   W^\prime_{\Phi }  = \qquad \nonumber \\  &=&
\bar{D}\bar{D}\bar{\Phi} \; {\cal K}^{\prime\prime}_{\Phi  \bar{\Phi}}(\Phi , \bar{\Phi}) +
  \bar{D}_{\dot\alpha}\bar{\Phi} \, \bar{D}^{\dot\alpha}\bar{\Phi} \; {\cal K}^{\prime\prime\prime}_{\Phi \bar{\Phi}\bar{\Phi}}(\Phi , \bar{\Phi}) + W^\prime_{\Phi }(\Phi ) =0 \; ,
 \qquad
\end{eqnarray}
where prime  denotes the derivative with respect to argument,  ${\cal K}^\prime_{\bar{\Phi}} := {\partial {\cal K}({\Phi},\bar{\Phi})\over \partial {\bar{\Phi}}}$, $\;{\cal K}^{\prime\prime}_{\bar{\Phi}\bar{\Phi}} := {\partial^2 \over \partial {\bar{\Phi}}^2}{\cal K}({\Phi},\bar{\Phi})$, {\it etc.} These equations can be obtained by solving the chirality conditions (\ref{bDP=0}) and (\ref{DbP=0}) in terms of prepotential, generic complex superfield $P$ ($=(\bar{P})^*$),
\begin{eqnarray}\label{Phi=bDbDP}
\Phi=\bar{D}\bar{D}P\; , \qquad \bar{\Phi}=DD\bar{P}\; , \qquad
\end{eqnarray}
and vary with respect to this prepotential and its complex conjugate,
\begin{eqnarray}\label{EqM=S/P}
{{\cal E}}= {\delta S_{s-int}[\Phi ; \bar{\Phi}]\over \delta {P}}\; , \qquad \bar{{\cal E}}= {\delta S_{s-int}[\Phi ; \bar{\Phi}]\over\delta \bar{P}}\; .  \qquad
\end{eqnarray}

\bigskip

\subsection{Four form field strength constructed from  the scalar  supermultiplet}

Having a chiral superfield $K$,
\begin{eqnarray}\label{bDK=0}
\bar{D}_{\dot{\alpha}}K =0 \; , \qquad {D}_{{\alpha}}\bar{K} =0 \; , \qquad
\end{eqnarray}
one can construct the following supersymmetric invariant closed four
form (CE cocycle) in flat $D=4$, ${\cal N}=1$ superspace
\cite{Gates:1980ay}
\begin{eqnarray}\label{F4:=dC3=}
F_4=dC_3 :={1\over 4} E^b\wedge E^a \wedge E^\alpha\wedge E^\beta \sigma_{ab}{}_{\alpha\beta} \; \bar{K} + {1\over 4} E^b\wedge E^a \wedge \bar{E}^{\dot{\alpha}}\wedge \bar{E}^{\dot{\beta}} \tilde{\sigma}_{ab}{}_{\dot{\alpha}\dot{\beta}}\; K + \qquad \nonumber \\
+ {1\over 4!} E^c\wedge E^b \wedge E^a \wedge E^\alpha\epsilon_{abcd} \sigma^d_{\alpha\dot{\beta}} \;\bar{D}{}^{\dot{\beta}} \bar{K} + {1\over 4!} E^c\wedge E^b \wedge E^a \wedge \bar{E}{}^{\dot{\beta}}\epsilon_{abcd} \sigma^d_{\alpha\dot{\beta}} \;{D}{}^{\alpha} {K} - \qquad \nonumber \\
 + {1\over 4!}  E^d \wedge E^c\wedge E^b \wedge E^a {i\over 4}\epsilon_{abcd}\left(\bar{D}\bar{D}\bar{K}- DDK\right) .  \qquad
\end{eqnarray}
Notice that we intentionally have not used the notation $\Phi$ for
chiral superfield to stress that, {\it e.g.} having a free chiral
superfield of Eqs. (\ref{bDP=0}) satisfying equations of motion
(\ref{DDP=0}), one can construct the three form using some holomorphic
functions $K=K(\Phi)$, $\bar{K}=\bar{K}(\overline{\Phi})$ which obey
$DDK= K^{\prime\prime}(\Phi) \, D^\alpha \Phi  \, D_\alpha \Phi  $
instead of (\ref{DDP=0}).

Interestingly enough, $F_4$ in (\ref{F4:=dC3=}) can be considered as real part of the complex closed
form  ${\cal F}^{L}_4$, \begin{eqnarray}\label{F4=cF4L+cF4R}
& {F}_4= \Re e ({\cal F}^{L}_4):= {1\over 2}
\left({\cal F}^{L}_4 + {\cal F}^{R}_4\right)\; ,   \qquad
\\ \label{cF4L=ch}
& {\cal F}^L_4 = {1\over  4 }E^b\wedge E^a \wedge  E^\alpha \wedge E^\beta \;
\sigma_{ab \; \alpha\beta}\, \bar{K}\; +  {1\over  4!} E^c\wedge E^b\wedge E^a \wedge  E^\alpha \;
\epsilon_{abcd} \sigma^d_{\alpha\dot{\beta}}\, \bar{D}{}^{\dot{\beta}}\bar{K}\;  + \nonumber \\ & +
{1\over  4!} E^d\wedge E^c\wedge E^b\wedge E^a \;
{i\over 16} \epsilon_{abcd} \, \bar{D}\bar{D}\bar{K}\; , \; \qquad {D}_\alpha\bar{K}=0\; , \; \qquad
\\ \label{cF4R=ch}
& {\cal F}^R_4 = {1\over  4 }E^b\wedge E^a \wedge  {\bar{E}}{}^{\dot{\alpha}} \wedge {\bar{E}}{}^{\dot{\beta}} \;
\tilde{\sigma}_{ab \; \dot{\alpha}\dot{\beta}}\, {K}\; +
{1\over  4!} E^c\wedge E^b\wedge E^a \wedge  {\bar{E}}{}^{\dot{\alpha}} \;
\epsilon_{abcd} \sigma^d_{{\beta}\dot{\alpha}}\, {D}{}^{{\beta}}{K}\;  + \nonumber \\ & +
{1\over  4!} E^d\wedge E^c\wedge E^b\wedge E^a \;
{i\over 16}\epsilon_{abcd} \, {D}{D}{K}\; , \; \qquad \bar{D}_{\dot{\alpha}}{K}=0\; . \qquad
\end{eqnarray}
The fact that these forms are closed as a consequences of (\ref{bDK=0}),
\begin{eqnarray}\label{dcF4L=0}
 d{\cal F}^L_4 = 0 \qquad &\Leftrightarrow & \qquad {D}_A\bar{K}=0\; , \; \qquad
\\ \label{dcF4R=0}
d{\cal F}^R_4 = 0 \qquad &\Leftrightarrow & \qquad  \bar{D}_{\dot{A}}{K}=0\; , \qquad
\end{eqnarray}
suggests the existence of the complex 3-form potentials  $C^L_3$ and ${C}^R_3=(C^L_3)^*$ such that
${\cal F}^L_4 =dC^L_3$ and ${\cal F}^R_4 =dC^R_3$.

To study a  supermembrane in the background of scalar multiplet, which will be the subject of the next section,  we do not need an explicit expression for $C_3$. However, we do need it to obtain the equations for the scalar multiplet fields with a source from supermembrane, so that we will come back to discussing the problem of constructing potentials in Sec. \ref{IntEqs}.

\bigskip

\section{$D=4$, ${\cal N}=1$ supermembrane and its interaction with scalar multiplet background}
\setcounter{equation}{0}

\subsection{Free supermembrane in flat $D=4$ ${\cal N}=1$ superspace}

The action for a free supermembrane in $D=4$, ${\cal N}=1$  superspace reads \cite{4Dsuperm}
\begin{eqnarray}\label{SM2=4D}
 S_{p=2} &=&  {1\over 2}\int d^3 \xi \sqrt{ g} -   \int\limits_{W^3} \hat{c}_3 = \; \qquad \nonumber \\ &=&   -{1\over 6}\int\limits_{W^3}  *\hat{E}_a\wedge \hat{E}^a  -   \int\limits_{W^3} \hat{c}_3\;   ,  \qquad
\qquad
\end{eqnarray}
where, in the first line $g= det(g_{mn})$ is the determinant of the induced metric,
\begin{eqnarray}\label{g=EE}
g_{mn}= \hat{E}_m^{a}\eta_{ab} \hat{E}_n^{b}\; , \qquad
\hat{E}_m^{a}:= \partial_m \hat{x}^a - i \partial_m\hat{\theta}{}^{\alpha} \sigma^a_{\alpha\dot{\alpha}} \hat{\bar{\theta}}{}^{\dot{\alpha}} + i\hat{\theta}{}^{\alpha} \sigma^a_{\alpha\dot{\alpha}} \partial_m\hat{\bar{\theta}}{}^{\dot{\alpha}}\; , \qquad
\end{eqnarray}
$W^3$ is the supermembrane worldvolume the embedding of which into the target superspace $\Sigma^{(4|4)}$ is defined parametrically by the coordinate functions
$\hat{z}{}^{{ {M}}}(\xi)= (\hat{x}{}^{a}(\xi)\, ,
\hat{\theta}^{{\alpha}}(\xi), \hat{\bar{\theta}}{}^{\dot{{\alpha}}}(\xi))$; $\xi^m= (\xi^0, \xi^1, \xi^2)$ are local coordinates on $W^3$,
\begin{eqnarray}\label{W3inS44}
W^3\; \subset  \Sigma^{(4|4)}\; : \qquad z^M= \hat{z}{}^{{ {M}}}(\xi)= (\hat{x}{}^{a}(\xi)\, ,
\hat{\theta}^{{\alpha}}(\xi), \hat{\bar{\theta}}{}^{\dot{{\alpha}}}(\xi))\; . \qquad
\end{eqnarray}
Finally,
\begin{eqnarray}\label{hc3=4D}
\hat{c}_3:= {1\over 3!} \hat{E}^{a_3} \wedge \hat{E}^{a_2} \wedge   \hat{E}^{a_1}
c_{a_1a_2a_3}(\hat{Z}) =
{1\over 3!}
d\xi^{m_3} \wedge
d\xi^{m_2} \wedge
d\xi^{m_1} \hat{c}_{m_1m_2m_3}= \qquad \nonumber \\ = -{1\over 6} d^3\xi \epsilon^{m_1m_2m_3}\hat{c}_{m_1m_2m_3} \;  \qquad
\end{eqnarray}
is the pull--back of the 3-form defined in Eq. (\ref{h4:=C-E}) to $W^3$, so that  the second, Wess--Zumino part of the action can be written in the form of (see \cite{4Dsuperm})  $\int\limits_{W^3} \hat{c}_3= -{1\over 6}\int d^3\xi \epsilon^{m_1m_2m_3}\hat{c}_{m_1m_2m_3}$.

In this paper we consider the case of closed supermembrane so that the worldvolume $W^3$ has no boundary, $\partial W^3= 0\!\!\! /$, and $\int_{W^3}d(...)=0$.
Then we do not need in the explicit form of $\hat{c}_{m_1m_2m_3}$ in (\ref{hc3=4D}) as far as variation of its integral in (\ref{SM2=4D}) can be calculated (using the Lie derivative formula, $\delta c_3= i_\delta dc_3+ di_\delta c_3$) through its exterior derivative, the pull--back $\hat{h}_4:= h_4(\hat{Z})$ of the CE cocycle (\ref{h4:=C-E}), $h_4=dc_3$.

In the second line of Eq. (\ref{SM2=4D}) we have written the first, Nambu-Goto term of the action as an integral of a differential three form. This is constructed from the pull--back of the bosonic vielbein form
\begin{eqnarray}\label{hEa=dxiE}
\hat{E}{}^a= d\xi^m \hat{E}_m^{a}\; , \qquad
\hat{E}_m^{a}:= \partial_m \hat{x}^a - i \partial_m\hat{\theta}{}^{\alpha} \sigma^a_{\alpha\dot{\alpha}} \hat{\bar{\theta}}{}^{\dot{\alpha}} + i\hat{\theta}{}^{\alpha} \sigma^a_{\alpha\dot{\alpha}} \partial_m\hat{\bar{\theta}}{}^{\dot{\alpha}}\; , \qquad
\end{eqnarray}
using the worldvolume Hodge star operation,
\begin{eqnarray}\label{*Ea:=}
*\hat{E}^a:= {1\over 2}d\xi^m\wedge d\xi^n\sqrt{g}\epsilon_{mnk}g^{kl}\hat{E}_l^a \; . \qquad
\end{eqnarray}

The action (\ref{SM2=4D}) is invariant under the local fermionic $\kappa$--symmetry transformations. These have the form of
\begin{eqnarray}\label{kappaX=}
\delta_\kappa x^\mu =
i \kappa^{\alpha} \sigma^\mu_{\alpha\dot{\alpha}} \bar{\theta}{}^{\dot{\alpha}} - i\theta^{\alpha} \sigma^\mu_{\alpha\dot{\alpha}}  \bar{\kappa}{}^{\dot{\alpha}} \; ,  \qquad \delta_\kappa \theta^{\alpha}= \kappa^{\alpha}\; , \qquad \delta_\kappa \bar{\theta}{}^{\dot{\alpha}} = \bar{\kappa}{}^{\dot{\alpha}} \; , \qquad
\end{eqnarray}
where the spinorial fermionic parameter $\kappa^\alpha =\kappa^\alpha (\xi) = (\bar{\kappa}{}^{\dot{\alpha}})^*$ has actually only two independent components because  it obeys the equations
\begin{eqnarray}\label{bk=kbg}
 \bar{\kappa}_{\dot{\alpha}}= \kappa^\beta {\bar{\gamma}}_{\beta\dot{\alpha}}\qquad \Leftrightarrow \qquad
 \kappa^\alpha= \bar{\kappa}_{\dot{\alpha}}
 \tilde{\bar{\gamma}}{}^{\dot{\beta}\alpha} \;  \quad
\end{eqnarray}
with
\begin{eqnarray}\label{bg=}
 {\bar{\gamma}}_{\beta\dot{\alpha}}= \epsilon_{\beta\alpha} \epsilon_{\dot{\alpha}\dot{\beta}}
 \tilde{\bar{\gamma}}{}^{\dot{\beta}\alpha}  = {i\over 3! \sqrt{g}}\sigma^a_{\beta\dot{\alpha}}\epsilon_{abcd}
 \epsilon^{mnk} \hat{E}_m^b\hat{E}_n^c\hat{E}_k^d\;  . \qquad
\end{eqnarray}
By construction, the matrix ${\bar{\gamma}}$ obeys
\begin{eqnarray}\label{gbg=I}
 {\bar{\gamma}}_{\beta\dot{\beta}}
 \tilde{\bar{\gamma}}{}^{\dot{\beta}\alpha}  = \delta_{\beta}
{}^{\alpha}\;  \qquad
\end{eqnarray}
which makes two equations in (\ref{bk=kbg}) equivalent.

To prove the $\kappa$--symmetry one have to use the identities
\begin{eqnarray}\label{EEEs=}
{1\over 2} \hat{E}{}^c \wedge \hat{E}{}^b \wedge \hat{E}{}^\alpha \sigma_{bc\alpha\beta}=
*\hat{E}_a\wedge \hat{E}{}^\alpha (\sigma^a\tilde{\bar{\gamma}})_{\alpha\beta}
\;  \qquad
\end{eqnarray}
which allows to present the variation of the kinetic, Nambu-Goto type, and the Wess--Zumino terms in similar form.

It is convenient to write the $\kappa$--symmetry transformations in the form of
\begin{eqnarray}\label{ikE0=}
i_\kappa \hat{E}{}^a := \delta_\kappa \hat{Z}^M E_M{}^a(\hat{Z}) =0 \; , \qquad \cases{ i_\kappa \hat{E}{}^\alpha := \delta_\kappa \hat{Z}^M E_M{}^\alpha (\hat{Z})
 = \kappa^\alpha= \bar{\kappa}_{\dot{\alpha}}
 \tilde{\bar{\gamma}}{}^{\dot{\beta}\alpha} \; , \cr i_\kappa \hat{\bar{E}}{}^{\dot\alpha}:= \delta_\kappa \hat{Z}^M E_M{}^{\dot{\alpha}}(\hat{Z})= \bar{\kappa}_{\dot{\alpha}}= \kappa^\beta {\bar{\gamma}}_{\beta\dot{\alpha}} \; .\qquad}
\end{eqnarray}

\bigskip

\subsection{Supermembrane action in the scalar multiplet background}

The action of supermembrane in the background of a scalar multiplet can be written in the form
\begin{eqnarray}\label{SM2+=4D}
S_{p=2} &=&  {1\over 2}\int d^3 \xi \sqrt{K\bar{K}}\sqrt{ g} -   \int\limits_{W^3} \hat{C}_3\;   ,  \qquad
\nonumber \\ &=&  -{1\over 6}\int\limits_{W^3}  *\hat{E}_a\wedge \hat{E}^a  \sqrt{K\bar{K}} -   \int\limits_{W^3} \hat{C}_3\;   ,  \qquad
\end{eqnarray}
where $\hat{C}_3$ is the pull--back of the $C_3$ potential defined by Eq. (\ref{F4:=dC3=}) involving the chiral superfields $K$ and $\bar{K}$ (\ref{bDK=0}). For simplify  we omit the hat symbol from the pull--backs of superfields here and below in the places where this cannot produce a confusion.

The action (\ref{SM2+=4D}) is invariant under the $\kappa$--symmetry
transformations
\begin{eqnarray}\label{ikE=}
i_\kappa \hat{E}{}^a =0 \; , \qquad i_\kappa \hat{E}{}^\alpha =
 \kappa^\alpha \; , \qquad i_\kappa \hat{\bar{E}}{}^{\dot\alpha}= \bar{\kappa}_{\dot{\alpha}} \; ,  \; \qquad
\end{eqnarray}
with $i_\kappa d\hat{Z}^M := \delta_\kappa \hat{Z}^M$, similar to ones in (\ref{ikE0=}) and (\ref{kappaX=}) but with the spinorial parameter obeying the reducibility conditions
\begin{eqnarray}\label{bk=Kkbg}
 \bar{\kappa}_{\dot{\alpha}}= - i  \kappa^\beta
 \bar{\gamma}_{\beta\dot{\alpha}}\; \sqrt{{\bar{K}/ K}}\; \qquad \Leftrightarrow \qquad
 \kappa^\alpha= i \bar{\kappa}_{\dot{\alpha}}
 \tilde{\bar{\gamma}}{}^{\dot{\beta}\alpha} \;  \sqrt{{K/{\bar{K}}}}\;  \qquad
\end{eqnarray}
defined by a projector which  differs from the one in (\ref{bk=kbg}) by a (super)field dependent phase factor
$ i \sqrt{{K^{}/{\bar{K}}}^{}}$.

Notice that, if we write the counterpart of the action (\ref{SM2+=4D}) with an arbitrary function ${\cal S}(K,\bar{K})$ instead of $ \sqrt{K\bar{K}}$ and perform the fermionic variation  (\ref{ikE=}) of such an action,  we find that the local fermionic  $\kappa$--symmetry parameter should obey the equations
$\kappa^\alpha \partial {\cal S}/\partial K = i/2 \bar{\kappa}_{\dot{\beta}}\tilde{\gamma}{}^{\dot{\beta}\alpha}$ and $\bar{\kappa}_{\dot{\beta}} \partial {\cal S}/\partial \bar{K} = \kappa^\alpha {\gamma}_{\alpha\dot{\beta}}$. This system of equations has a nontrivial solution when $\partial {\cal S}/\partial \bar{K} = { 1\over  4 \partial{\cal S}/\partial {K}}$. This latter equation is solved by ${\cal S}(K,\bar{K})=\sqrt{K\bar{K}}$ so that the action (\ref{SM2+=4D}) for scalar multiplet in supergravity background can be constructed from the requirement of the $\kappa$--symmetry.

\subsection{Equations of motion for supermembrane in a background of an off--shell scalar supermultiplet}

The supermembrane equations of motion can be obtained by varying the action (\ref{SM2+=4D}) with respect to coordinate functions $\hat{Z}{}^M(\xi)$, so that we can write them in the form of ${\delta S_{p=2}\over \delta \hat{Z}{}^M(\xi)}=0$. The convenient form of the bosonic and fermionic equations can be extracted by multiplying this on the inverse supervielbein, $E_\alpha^M(\hat{Z}){\delta S_{p=2}\over \delta \hat{Z}{}^M(\xi)}=0$ and $E_a^M(\hat{Z}){\delta S_{p=2}\over \delta \hat{Z}{}^M(\xi)}=0$. These combinations appear as the coefficients for $i_\delta \hat{E}^\alpha:=  \delta \hat{Z}{}^M(\xi)E_M{}^\alpha (\hat{Z})$, $i_\delta \hat{\bar{E}}{}^{\dot\alpha}:=  \delta \hat{Z}{}^M(\xi)\bar{E}_M{}^{\dot\alpha} (\hat{Z})$ and $i_\delta \hat{E}^a:=  \delta \hat{Z}{}^M(\xi)E_M{}^a (\hat{Z})$ in the integrand of the action variation. This implies the possibility to write the formal expression for supermembrane equations of motion in the form
\begin{eqnarray}\label{smEqm=f-gen}
  {\delta S_{p=2}/ i_\delta \hat{E}{}^\alpha }:= E_\alpha^M(\hat{Z}){\delta S_{p=2}\over \delta \hat{Z}{}^M(\xi)}=0\; &,&  \qquad  {\delta S_{p=2}/ i_\delta \hat{\bar{E}}{}^{\dot{\alpha}}}:= E_{\dot{\alpha}}^M(\hat{Z}){\delta S_{p=2}\over \delta \hat{Z}{}^M(\xi)}=0\; ,  \qquad \\ \label{smEqm=b-gen} {\delta S_{p=2}/ i_\delta \hat{E}{}^a }&:=& E_a^M(\hat{Z}){\delta S_{p=2}\over \delta \hat{Z}{}^M(\xi)}=0\; . \qquad
\end{eqnarray}
The straightforward calculation gives the following explicit form of these equations of motion
\begin{eqnarray}\label{smEqm=f}
&& *\hat{E}_a \wedge \left( i \hat{\bar{E}}{}^{\dot{\alpha}} \sigma^a_{\alpha\dot{\alpha}} \sqrt{K\bar{K}} - \hat{E}_\beta (\bar{\gamma}\tilde{\sigma}{}^a)_\alpha{}^\beta \bar{K} \right) + {1\over 12} *\hat{E}_a \wedge \hat{E}^a \left(\sqrt{\bar{K}/K} D_\alpha K- i\bar{\gamma}_{\alpha\dot{\alpha}} \bar{D}{}^{\dot{\alpha}}\bar{K} \right)=0\; ,  \nonumber \\ && \\ \label{smEqm=bf}
&& *\hat{E}_a \wedge \left( i \hat{{E}}{}^{{\alpha}} \sigma^a_{\alpha\dot{\alpha}} \sqrt{K\bar{K}} - \hat{\bar{E}}{}_{\dot\beta} (\tilde{\sigma}{}^a\bar{\gamma}){}^{\dot\beta}{}_{\dot\alpha} {K} \right) + {1\over 12} *\hat{E}_a \wedge \hat{E}^a \left(\sqrt{K/\bar{K}} \bar{D}_{\dot\alpha} \bar{K}- iD^\alpha K \bar{\gamma}_{\alpha\dot{\alpha}} \right)=0\; , \nonumber \\ && \end{eqnarray}
\begin{eqnarray} \label{smEqm=b} \; D(*\hat{E}_a) &- & {1\over 6} *\hat{E}_b \wedge \hat{E}{}^b \left({D_a\ln\bar{K}} +  {D_a\ln K} \right)+ {1\over 2}  *\hat{E}_a \wedge  ( d\ln \hat{\bar{K}} +  d\ln \hat{K}) - \qquad \nonumber \\ && - {i\over 12\sqrt{K\bar{K}}}\hat{E}{}^d\wedge\hat{E}{}^c\wedge  \hat{E}{}^b \,\epsilon_{abcd} \,  (\bar{D}\bar{D}\bar{K}-DDK) - \qquad \nonumber \\ && - {1\over 4\sqrt{K\bar{K}}}\hat{E}{}^c\wedge  \hat{E}{}^b \wedge \epsilon_{abcd} \sigma^d_{\alpha\dot{\alpha}}\,  (\hat{E}{}^\alpha \bar{D}{}^{\dot{\alpha}}\bar{K}+ \hat{\bar{E}}{}^{\dot{\alpha}}{D}{}^\alpha K) -  \qquad \nonumber \\  && - \hat{E}{}^b \wedge \hat{E}{}^\alpha \wedge \hat{E}{}^\beta \sigma_{ab}{}_{\alpha\beta}\; \sqrt{{\bar{K}}/{K}}  - \hat{E}{}^b \wedge \hat{\bar{E}}{}^{\dot\alpha} \wedge \hat{\bar{E}}{}^{\dot\beta} \tilde{\sigma}_{ab}{}_{\dot{\alpha}\dot{\beta}}\; \sqrt{K/{\bar{K}}}  = 0\; . \qquad
\end{eqnarray}

Notice that the above equations of motion are not independent. According to the second Noether theorem, the gauge symmetries of a dynamical system result in the so-called Noether identities relating the left-hand sides of equations of motion of this system. The supermembrane possesses a number of gauge symmetries, including the local fermionic $\kappa$--symmetry (\ref{ikE=}), (\ref{bk=Kkbg}). This is reflected by the fact that contracting our fermionic equation (\ref{smEqm=f}) with $i\sqrt{K/\bar{K}} \tilde{\bar{\gamma}}{}^{\dot{\beta}\alpha}$ we arrive at Eq. (\ref{smEqm=bf}).
Denoting the {\it l.h.s.}'s of Eqs. (\ref{smEqm=f}) and (\ref{smEqm=bf}) by $\Psi_\alpha$ and $\bar{\Psi}_{\dot\alpha}$, respectively, we can write the above described Noether identity for the  $\kappa$--symmetry in the form of
\begin{eqnarray} \label{NI=kappa}
\tilde{\bar{\gamma}}{}^{\dot{\beta}\alpha}\Psi_\alpha \equiv -i\sqrt{\bar{K}/K}\epsilon^{\dot{\beta\dot{\alpha}}} \bar{\Psi}_{\dot\alpha}\; . \qquad
\end{eqnarray}

\section{Superfield equations for the dynamical system of special scalar supermultiplet interacting with supermembrane}
\label{IntEqs}

\setcounter{equation}0

 \subsection{Special scalar multiplet and its dual three form potential}

In our discussion below we will be considering not generic but   {\it special scalar multiplet} described by the chiral superfield constructed from the  {\it real prepotential} $V=(V)^*$,
\begin{eqnarray}\label{Phi=DDV}
\Phi=\bar{D}\bar{D}V\; , \qquad \bar{\Phi}=DDV\;  . \qquad
\end{eqnarray}
On the level of auxiliary fields the distinction of this special case is that one of the real auxiliary scalars  of the generic scalar multiplet is replaced in it by a divergence of a real vector, $\partial_\mu k^\mu$ or, equivalently, by the field strength of a three form potential $k_{\nu\rho\sigma}= k^\mu \epsilon_{\mu\nu\rho\sigma}$ (in this latter form it was described in \cite{Gates:1980ay} and, as one of 'variant superfield representations', in \cite{Gates:1980az}).

Indeed, the complex prepotential $P$ of the generic chiral multiplet,  $\Phi=\bar{D}\bar{D}P$, $ \bar{\Phi}=DD\bar{P}$, is defined up to the gauge transformations, $\bar{P}\mapsto \bar{P} + D_\alpha \Xi^\alpha$. These imply that the imaginary part of the generic prepotential is transformed by
$\Im m\, P := (P-\bar{P})/2i\mapsto \Im  m\, P + (D_\alpha \Xi^\alpha - D_{\dot\alpha} \bar{\Xi}^{\dot\alpha})/2i$.
Hence not-pure gauge parts of the superfield parameter $\Im m\, P$ are the ones which do not have their exact counterparts in the composed superfield $(D_\alpha \Xi^\alpha - D_{\dot\alpha} \bar{\Xi}^{\dot\alpha})/2i$.
One can check that the superfield parameter $D_\alpha \Xi^\alpha - D_{\dot\alpha} \bar{\Xi}^{\dot\alpha}$ has all the components but one having contributions of different independent functions without derivatives. The only exception is the highest component in its decomposition  which reads $- 4i \theta\theta\, \bar{\theta}\bar{\theta} \partial_a(k^a+\bar{k}{}^a)$ and includes the divergence of the real part of the complex vector  $k_a=\tilde{\sigma}_a^{\dot{\alpha}\alpha}(\bar{D}_{\dot{\alpha}}\Xi_\alpha)\vert_{\theta=0}= (\bar{k}_a)^*$
versus an arbitrary function in a generic real scalar superfield, like $P$. Then one can guess that the equations of motion for the special scalar supermultiplet will differ from the set of equations for a generic scalar supermultiplet by that one of the algebraic auxiliary field equations of the latter ($({\cal E}-{\cal E})\vert_0=0$) will be replaced by its derivative ($\partial_a({\cal E}-{\cal E})\vert_0=0$). In other words, the general solution of the (auxiliary) field equations of the special scalar supermultiplet involves one additional (with respect to the generic case) arbitrary real constant. Furthermore, this indicates that the above mentioned auxiliary field equation of the special scalar supermultiplet ($\partial_a({\cal E}-{\cal E})\vert_0=0$) is dependent, {\it i.e.} can be obtained as a consequence of other equations; this implies that  the only effect of the use of the complex prepotential in the generic case is vanishing of a real constant which is indefinite in the case of special scalar multiplet (where $\partial_a({\cal E}-{\cal E})\vert_0=0\;\Rightarrow \; ({\cal E}-{\cal E})\vert_0=-2ic$). We will see that this is indeed the case.

 \subsubsection{Equations of motion of special scalar multiplet}

The variation of the general action (\ref{S=PhibPhi}) for the special chiral superfields  (\ref{Phi=DDV}) with respect to real prepotential $V$ apparently produces only the real part of the complex equation (\ref{EqM=sc0}),
\begin{eqnarray}\label{EqM=sc0+b}
{\delta S_{s-int}[\bar{D}\bar{D}V ; DDV]\over  \delta {V}}=0 \; \Rightarrow \qquad  {{\cal E}}+\bar{{\cal E}}&:= & \bar{D}\bar{D} {\cal K}^{\prime}_{{\Phi}} + DD {\cal K}^\prime_{\bar{\Phi}}  +  W^\prime_{\Phi } +  \bar{W}^\prime_{\bar{\Phi} }  = 0
\; .
 \qquad
\end{eqnarray}

However, as far as the left hand sides of the equations of motion for generic scalar multiplet, Eqs. (\ref{EqM=sc0}) and (\ref{EqM=sc0b}), are, respectively, anti-chiral and chiral, ${D}_{\alpha} \bar{{\cal E}}= 0$ and $\bar{D}_{\dot\alpha} {\cal E}=0$, Eq. (\ref{EqM=sc0+b}) implies that the imaginary part of the complex equation  (\ref{EqM=sc0}) is equal to a constant,
\begin{eqnarray}\label{EqM-bEqM=c}
{{\cal E}}+\bar{{\cal E}}=0 \; \qquad \Rightarrow \qquad  D_\alpha ({{\cal E}}-\bar{{\cal E}})=0\qquad \Rightarrow \qquad  \partial_a ({{\cal E}}-\bar{{\cal E}})=0 \; . \qquad
\end{eqnarray}
Hence the only effect of the use of the special chiral superfields (\ref{Phi=DDV}) instead of the generic scalar superfield (\ref{Phi=bDbDP}) is that the equation ${{\cal E}}=0$ is replaced by
${{\cal E}}=-ic$ where $c$ is an arbitrary real constant.

\bigskip

\subsubsection{On spontaneous supersymmetry breaking}

The presence of this arbitrary constant in the right hand side of the superfield equations of motion, ${{\cal E}}=-ic$,  actually suggests a possible spontaneous supersymmetry breaking in the theory of special chiral multiplet.
To clarify this, let us discuss the simple case of a free massless special scalar multiplet, in which the action reads
$\int d^8z \Phi\bar{\Phi}= \int d^8z DDV \bar{D}\bar{D}V$ so that ${\bar{\cal E}}=DD\Phi $ and the equations of motion (\ref{EqM=sc0+b}) simplify to
\begin{eqnarray}\label{EqM0=sc} DD\Phi  + \bar{D}\bar{D}\bar{\Phi}=0\; . \qquad
\end{eqnarray}
As it has been discussed above (Eq. (\ref{EqM-bEqM=c})) these equations lead  to $D_\alpha (DD\Phi  - \bar{D}\bar{D}\bar{\Phi}) =0$ and $\partial_a (DD\Phi  - \bar{D}\bar{D}\bar{\Phi}) =0$. Algebraically all this set of equations is solved by  $DD\Phi   =-ic$  with the above mentioned arbitrary real constant $c$,
\begin{eqnarray}\label{EqM=-ic} DD\Phi  + \bar{D}\bar{D}\bar{\Phi}=0 \qquad \Rightarrow \qquad
DD\Phi   =-ic\; , \qquad c=const\, . \qquad
\end{eqnarray}
In particular, this constant enters the solution of auxiliary field equations which now reads
\begin{eqnarray}\label{auxEq=-ic}
DD\Phi \vert_0  =-ic\; , \qquad c=const\, . \qquad
\end{eqnarray}
As the on-shell supersymmetry transformations of the fermionic fields $\psi_\alpha= -iD_\alpha\Phi\vert_0$ are obtained from the off-shell ones,
$\delta \psi_\alpha =  {i\over 2}\varepsilon_\beta\, DD\Phi\vert_0 + 2(\sigma^a\bar{\varepsilon})_\alpha \partial_a \phi$, by inserting the above solution  of the auxiliary field equations, they read \begin{eqnarray}\label{susy=ic}\delta \psi_\alpha = {c\over 2} \, \varepsilon_\beta\, + 2(\sigma^a\bar{\varepsilon})_\alpha \partial_a \phi\, . \qquad
\end{eqnarray}
Hence, for nonvanishing value of $c$, the on-shell supersymmetry transformations of $\delta \psi_\alpha$ contains the additive contribution of supersymmetry parameter $\varepsilon_\beta\,$ characteristic of the transformation rules of the Volkov-Akulov Goldstone fermion \cite{VA72} the presence of which may be considered as an indication of the spontaneous supersymmetry breaking.

However, studying more carefully the case of free special scalar multiplet, one finds that such a spontaneous symmetry breaking actually does not occur if nontrivial boundary conditions are not introduced\footnote{One of the authors (I.B.) is grateful to Warren Siegel for the discussion on this issue.}. Indeed, the constant in the superfield equations (\ref{EqM=-ic}) can be reproduced from the generic scalar supermultiplet action which includes the superpotential linear in chiral superfield, $W(\Phi)=-ic\Phi$. As it was observed already in \cite{Zumino+Iliopoulos=74},  such a term can be removed from the action by a field redefinition. However, the boundary term contribution may change the situation; this role can be also played by supermembrane contribution. We leave the discussion on spontaneous supersymmetry breaking in the interacting system of scalar multiplet and supermembrane for future and now turn to the three form potential presentation of the special chiral supermultiplet.

\subsubsection{Dual three form potential}

The four form field strength constructed with the use of special scalar multiplet (\ref{Phi=DDV}) is obtained from (\ref{F4:=dC3=}) by substituting
\begin{eqnarray}\label{K=DDV}
\bar{K}=\bar{\Phi}=DDV\;  , \qquad K= {\Phi}= \bar{D}\bar{D}V\; . \qquad
\end{eqnarray}
It reads
\begin{eqnarray}\label{F4=F4(V)}
F_4 = dC^\prime_3 &= {1\over  4 }E^b\wedge E^a \wedge  E^\alpha \wedge E^\beta \;
\sigma_{ab \, \alpha\beta} DDV\, + {1\over  4}E^b\wedge E^a \wedge  \bar{E}{}^{\dot{\alpha}} \wedge \bar{E}{}^{\dot{\beta}} \;
\tilde{\sigma}_{ab \, \dot{\alpha}\dot{\beta}} \bar{D}\bar{D}V\, + \nonumber \\ & +
{1\over  4!} E^c\wedge E^b\wedge E^a \wedge \left( E^\alpha \;
\epsilon_{abcd} \sigma^d_{{\alpha}\dot{\beta}}\, \bar{D}{}^{\dot{\beta}}DDV\; +   \bar{E}{}^{\dot{\alpha}} \;
\epsilon_{abcd} \sigma^d_{{\beta}\dot{\alpha}}\, {D}{}^{{\beta}}\bar{D}\bar{D}V\;\right) \;  + \nonumber \\ & +
{1\over  4!} E^d\wedge E^c\wedge E^b\wedge E^a \;
{i\over 4} \epsilon_{abcd} \, (\bar{D}\bar{D}DDV - DD\bar{D}\bar{D}V)\; . \; \qquad
\end{eqnarray}
The corresponding 3-form potential $C^\prime_3$ can be written in terms of the real prepotential as follows
\cite{Gates:1980ay}
\begin{eqnarray}\label{C3'=}
 C^\prime_3 &=& 2i E^c \wedge  E^\alpha \wedge \bar{E}{}^{\dot{\beta}} \; {\sigma}_{c \; {\alpha}\dot{\beta}}\; V+  {1\over 2}E^c\wedge E^b \wedge  E{}^{{\alpha}}  \sigma_{bc \; {\alpha}}{}^{\beta}\, D_\beta V\; - \nonumber \\ &&
 -  {1\over  2}E^b\wedge E^a \wedge  \bar{E}{}^{\dot{\beta}} \;
\tilde{\sigma}_{ab}{}^{\dot{\alpha}}{}_{\dot{\beta}}\, \bar{D}_{\dot{\alpha}}V\; -
 {1\over 4!} E^c\wedge E^b\wedge E^a
 \epsilon_{abcd}  \tilde{\sigma}^{d \; \dot{\beta}\alpha} \,  [{D}_\alpha \, , \, \bar{D}_{\dot{\beta}}] V \; . \; \qquad
\end{eqnarray}
Of course, this expression can be changed on an equivalent one using gauge transformations
$\delta C_3=d\alpha_2$. These do not change the field strength (\ref{F4=F4(V)}) and are responsible for the possibility to do not have the lower dimensional form contributions ($\propto E^\alpha\wedge E^\beta \wedge E^\gamma$ {\it etc.}) in the above $C_3^\prime$.

The existence  of this simple three form $C_3^\prime$ giving a dual description of the special chiral supermultiplet (\ref{Phi=DDV}) is the main reason to restrict our discussion below by this special case.

\subsection{Superfield equations of motion for interacting system}

Let us consider the most general interaction of the special scalar supermultiplet with supermembrane
as described by the action  (\ref{SM2+=4D}) with $K= \Phi= \bar{D}_{\dot{\alpha}}\bar{D}{}^{\dot{\alpha}}V$ and
$\bar{K}=  \bar{\Phi }= {D}^{{\alpha}}{D}_{{\alpha}}V $ as in (\ref{K=DDV}), {\it i.e.} by
\begin{eqnarray}\label{Sm+SM2=4D}
S&=& \int d^8z\, {\cal K}(\Phi , \bar{\Phi}) +  \int d^6\zeta_L \, W(\Phi )  + c.c.
+  {1\over 2}\int d^3 \xi \, \sqrt{ g}\, \sqrt{\hat{\Phi}\hat{\bar{\Phi}}} -   \int\limits_{W^3} \hat{C}^\prime_3= \qquad
 \\ &=& \int d^4x\, \bar{D}\bar{D}DD{\cal K}(\Phi , \bar{\Phi}) +  \int d^4x (DDW(\Phi )  + c.c.)   -{1\over 6}\int\limits_{W^3}  *\hat{E}_a\wedge \hat{E}^a  \sqrt{\hat{\Phi}\hat{\bar{\Phi}}} -   \int\limits_{W^3} \hat{C}_3^\prime\; \nonumber
\end{eqnarray}
with special chiral superfield (\ref{Phi=DDV}),
\begin{eqnarray}\label{Phi=DDV2}
 \Phi= \bar{D}_{\dot{\alpha}}\bar{D}{}^{\dot{\alpha}}V \; ,  \qquad   \bar{\Phi }= {D}^{{\alpha}}{D}_{{\alpha}}V \; .  \qquad
\end{eqnarray}

The variation of the interacting action (\ref{Sm+SM2=4D}) with respect to supermembrane variables
gives formally the same equations of motion as for the supermembrane in the background, (\ref{smEqm=f})--(\ref{smEqm=b}), but with $K=\Phi=DDV$, \footnote{As above, to simplify the expressions, we omitted the hat symbol form the pull--backs of superfields and their derivatives in Eqs. (\ref{smEqm=bfs}) and (\ref{smEqm=bs}), but, in contrast, left all the pull--back symbols in Eq. (\ref{smEqm=fs}) so that one can appreciate simplification comparing this with its complex conjugate Eq. (\ref{smEqm=bs}).}
 \begin{eqnarray}\label{smEqm=fs}
&& *\hat{E}_a \wedge \left( i \hat{\bar{E}}{}^{\dot{\alpha}} \sigma^a_{\alpha\dot{\alpha}} \sqrt{\hat{\Phi}\hat{\bar{\Phi}}} - \hat{E}_\beta (\bar{\gamma}\tilde{\sigma}{}^a)_\alpha{}^\beta \hat{\bar{\Phi}} \right) + {1\over 12} *\hat{E}_a \wedge \hat{E}^a \left(\sqrt{\hat{\bar{\Phi}}/\hat{\Phi}} \, \widehat{D_\alpha \Phi }- i\bar{\gamma}_{\alpha\dot{\alpha}} \widehat{\bar{D}{}^{\dot{\alpha}}\bar{\Phi}} \right)=0\; ,  \nonumber \\ && \\ \label{smEqm=bfs}
&& *\hat{E}_a \wedge \left( i \hat{{E}}{}^{{\alpha}} \sigma^a_{\alpha\dot{\alpha}} \sqrt{\Phi\bar{\Phi}} - \hat{\bar{E}}{}_{\dot\beta} (\tilde{\sigma}{}^a\bar{\gamma}){}^{\dot\beta}{}_{\dot\alpha} {\Phi} \right) + {1\over 12} *\hat{E}_a \wedge \hat{E}^a \left(\sqrt{\Phi/\bar{\Phi}} \bar{D}_{\dot\alpha} \bar{\Phi}- iD^\alpha \Phi \bar{\gamma}_{\alpha\dot{\alpha}} \right)=0\; , \nonumber \\ && \end{eqnarray}
\begin{eqnarray} \label{smEqm=bs} \; D(*\hat{E}_a) &- & {1\over 6} *\hat{E}_b \wedge \hat{E}{}^b \left({D_a\ln\bar{\Phi}} +  {D_a\ln \Phi} \right)+ {1\over 2}*\hat{E}_a \wedge  ( d\ln \hat{\bar{\Phi}} +  d\ln \hat{\Phi}) - \qquad \nonumber \\ && - {i\over 12\sqrt{\Phi\bar{\Phi}}}\hat{E}{}^d\wedge\hat{E}{}^c\wedge  \hat{E}{}^b \,\epsilon_{abcd} \,  (\bar{D}\bar{D}\bar{\Phi}-DD\Phi) - \qquad \nonumber \\ && - {1\over 4\sqrt{\Phi\bar{\Phi}}}\hat{E}{}^c\wedge  \hat{E}{}^b \wedge \epsilon_{abcd} \sigma^d_{\alpha\dot{\alpha}}\,  (\hat{E}{}^\alpha \bar{D}{}^{\dot{\alpha}}\bar{\Phi}+ \hat{\bar{E}}{}^{\dot{\alpha}}{D}{}^\alpha \Phi) -  \qquad \nonumber \\  && - \hat{E}{}^b \wedge \hat{E}{}^\alpha \wedge \hat{E}{}^\beta \sigma_{ab}{}_{\alpha\beta}\; \sqrt{{\bar{\Phi}}/{\Phi}}  - \hat{E}{}^b \wedge \hat{\bar{E}}{}^{\dot\alpha} \wedge \hat{\bar{E}}{}^{\dot\beta} \tilde{\sigma}_{ab}{}_{\dot{\alpha}\dot{\beta}}\; \sqrt{\Phi/{\bar{\Phi}}}  = 0\; . \qquad
\end{eqnarray}
However, the target superspace superfields the pull--backs of which enter these equations have to be the solutions of interacting equations with the source terms from the supermembrane. These superfield interacting equations read
\begin{eqnarray}\label{Eqm=J}
 {\cal E}+\bar{\cal E} = J\; ,  \qquad
\end{eqnarray}
where
\begin{eqnarray}\label{Eqm=J}
 {\cal E}=
 \bar{D}\bar{D}\bar{\Phi} \; {\cal K}^{\prime\prime}_{\Phi  \bar{\Phi}}(\Phi , \bar{\Phi}) +
  \bar{D}_{\dot\alpha}\bar{\Phi} \, \bar{D}^{\dot\alpha}\bar{\Phi} \; {\cal K}^{\prime\prime\prime}_{\Phi \bar{\Phi}\bar{\Phi}}(\Phi , \bar{\Phi}) + W^\prime_{\Phi }(\Phi )\;  \nonumber \\
\end{eqnarray}
(see  (\ref{EqM=sc0}) and (\ref{EqM=sc0+b}))  and
\begin{eqnarray}\label{J=dS/dV}
J(z) = - {\delta S_{p=2}\over \delta V(z)}
\;   \qquad
\end{eqnarray}
is the current superfield from the supermembrane. The problem of obtaining the complete set of interacting equations for the dynamical system of supermembrane and special chiral supermultiplet is now reduced to the problem of calculating this supermembrane current.

\subsection{Supermembrane current}
The supermembrane current is split naturally on the contributions from the Nambu--Goto and the Wess--Zumino terms of the action (\ref{SM2+=4D}) with (\ref{Phi=DDV})
\begin{eqnarray}\label{J=JNG+JWZ}
J(z) = J^{NG}(z)+ J^{WZ}(z) = - {\delta S_{p=2}\over \delta V(z)}
\;   \qquad
\end{eqnarray}
The Nambu-Goto part of the current
\begin{eqnarray}\label{JNG:=}
 & J^{NG}(z) := - {\delta \over \delta V(z)}  {1\over 2}\int d^3 \xi \, \sqrt{ g} \sqrt{\widehat{\bar{D}\bar{D}V}\, \widehat{DDV}}
\end{eqnarray}
($\widehat{DDV}:= D^\alpha D_\alpha V(z)\vert_{z^M=\hat{z}{}^M(\xi)}$)
is calculated by first using the properties of the superspace delta function
\begin{eqnarray}\label{d8:=}
 \delta^8(z):={1\over 16}\,\delta^4(x) \, \theta\theta \, \bar{\theta}\bar{\theta}\; , \qquad
 \int d^8z\, \delta^8(z-z') f(z) = f(z') \;
\end{eqnarray}
to present (\ref{JNG:=}) in the form
\begin{eqnarray}\label{JNG=intd8z}
 & J^{NG}(z) = - {\delta \over \delta V(Z)}  {1\over 2} \int d^8z'\, \sqrt{{\bar{D}\bar{D}V(z')}\, {DDV}(z')} \, \int d^3 \xi \, \sqrt{ g} \delta^8(z'-\hat{z})\; . \qquad
\end{eqnarray}
Then the calculation reduces to using the definition of variation
${\delta V(z')\over \delta V(z)}=\delta^8(z'-{z})$ and performing the superspace integration. In such a way one arrives at
\begin{eqnarray}\label{JNG=}
 J^{NG}(Z) &=& - {1\over 4} \, \int d^3 \xi \, \sqrt{g} \, \sqrt{{\hat{\Phi}\over \hat{\bar{\Phi}}}}\; {D}{D} \delta^8(z-\hat{z})  - {1\over 4} \, \int d^3 \xi \, \sqrt{g}\, \sqrt{{\hat{\bar{\Phi}}\over\hat{ \Phi}}}\; \bar{D}\bar{D}\delta^8(z-\hat{z}) \, , \qquad
\end{eqnarray}
where
$\Phi=\bar{D}\bar{D}V$, $\bar{\Phi}=DDV$ (see Eqs. (\ref{Phi=bDbDP})) and
$\hat{\Phi}:= {\Phi}(\hat{z}(\xi))$ {\it etc.}
Similarly one can present   the Wess--Zumino current in the form of
\begin{eqnarray}\label{JWZ=}
J^{WZ}(Z) &=& \left(2i \int_{W^3} \hat{E}^c\wedge \hat{E}{}^\alpha \wedge \hat{E}{}^{\dot\alpha} \sigma_{c\alpha\dot\alpha}  + \right. \nonumber \\ && \left. + {1\over 2}  \int_{W^3} \hat{E}^c\wedge \hat{E}^b\wedge \hat{E}{}^\alpha \sigma_{bc\alpha}{}^{\beta}D_\beta  -  {1\over 2}  \int_{W^3} \hat{E}^c\wedge \hat{E}^b \wedge \hat{E}{}^{\dot\alpha} \tilde{\sigma}_{bc}{}^{\dot{\beta}}{}_{\dot\alpha} \bar{D}_{\dot{\beta}} - \right. \nonumber \\ && \left. - {1\over 4!}  \int_{W^3} \hat{E}^c\wedge \hat{E}^b \wedge \hat{E}^a\epsilon_{abcd} \tilde{\sigma}^{d\dot{\alpha}\alpha} [D_\alpha, \bar{D}_{\dot{\alpha}}] \right)  \delta^8(z-\hat{z})\; . \qquad
\end{eqnarray}

\section{Simplest equations of motion for spacetime fields interacting with dynamical supermembrane}
\setcounter{equation}0

Having the superfield equations with supermembrane current contributions, the next stage is to extract the equations of motion for the physical fields of the supermultiplet. In this paper we will do this for the  simplest case when the special scalar multiplet part of the interacting action is given by
the kinetic term (\ref{S=PhibPhi}) only, this is to say for the interacting system described by the action
\begin{eqnarray}\label{S=S0+SM2}
S=S_{kin}+ S_{p=2} = \int d^8z\, \Phi \bar{\Phi}  + {1\over 2}\int d^3 \xi \sqrt{\hat{\Phi}\hat{\bar{\Phi}}}\sqrt{ g} -   \int\limits_{W^3} \hat{C}_3{}^\prime\;     \qquad
\end{eqnarray}
where $\Phi=\bar{D}\bar{D}V$,  $\bar{\Phi}=DDV$ (\ref{Phi=DDV2}) and $ \hat{C}^\prime_3$ is the pull--back to $W^{3}$ of the 3-form ${C}^\prime_3$ defined in (\ref{C3'=}). The interacting equations of motion for the bulk superfields, Eqs.  (\ref{Eqm=J}), in this case simplifies to
\begin{eqnarray}\label{Eqm0=J}
  DD\Phi+\bar{D}\bar{D}\bar{\Phi} = J(z)\;   \qquad
\end{eqnarray}
where the current $J(z)$ is given by (\ref{J=JNG+JWZ}), (\ref{JNG=}) and (\ref{JWZ=}).

\subsection{General structure of the simplest special scalar multiplet equations with a superfield source}

Superfield equation (\ref{Eqm0=J}) encodes the dynamical equations for the physical fields of the scalar multiplet, $\phi(x)=\Phi\vert_0$ and $\psi_\alpha (x)=- i(D_\alpha\Phi )\vert_0$, as well as algebraic equations for auxiliary fields  $DD\Phi\vert_0$ and $\bar{D}\bar{D}\bar{\Phi}\vert_0 $. These latter include the leading component of the real superfield equation (\ref{Eqm0=J})
\begin{eqnarray}\label{Eqm0=Jv0}
 DD\Phi \vert_0 +\bar{D}\bar{D}\bar{\Phi} \vert_0 = J(z)\vert_0\;   \qquad
\end{eqnarray}
as well as the first order equation
\begin{eqnarray}\label{d(aux-aux)=}
 \partial_a( DD\Phi \vert_0 -\bar{D}\bar{D}\bar{\Phi} \vert_0 ) = - {i\over 4} \tilde{\sigma}_a^{\dot{\alpha}\alpha}[D_\alpha \, , \, \bar{D}_{\dot\alpha}] J(z)\vert_0\;  . \qquad
\end{eqnarray}
The set of dynamical field equations include the Dirac (actually Weyl) equation with the source from supermembrane,
\begin{eqnarray}\label{Dirac=J}
{\sigma}^a_{\alpha\dot{\alpha}}\partial_a\psi^\alpha:= - i \partial_{\alpha\dot{\alpha}}D_\alpha\Phi \vert_0 = {1\over 4} \bar{D}_{\dot{\alpha}}J(z)\vert_0 \; , \qquad
\end{eqnarray}
and the Klein-Gordon equation, also with the source,
\begin{eqnarray}\label{K-G=J}
\Box \phi(x):= \Box \Phi\vert_0 = -{1\over 16} \bar{D}\bar{D}J(z)\vert_0 \;  . \qquad
\end{eqnarray}
Now, to specify supermembrane contributions to the scalar multiplet field equations we have to calculate the derivatives of the supermembrane current.

\subsection{Dynamical scalar multiplet equations with supermembrane source contributions}

\subsubsection{Auxiliary field equations}

The leading components $J\vert_0$  of the current $J$ in (\ref{Eqm0=Jv0}) is the sum of
\begin{eqnarray}\label{JNG0=}
 J^{NG}\vert_{0} &=&  {1\over 16} \,\sqrt{{\phi\over \bar{\phi}}}\; \int d^3 \xi \, \sqrt{g} \, \hat{\bar{\theta}}\hat{\bar{\theta}} \delta^4(x-\hat{x}) + {1\over 16} \sqrt{{\bar{\phi}\over \phi}}\; \int d^3 \xi \, \sqrt{g}\, \hat{{\theta}}\hat{{\theta}}\delta^4(x-\hat{x})  \;  \qquad
\end{eqnarray}
and
\begin{eqnarray}\label{JWZ0=}
J^{WZ}(Z)\vert_{0}  &=& {1\over 48}  \int_{W^3} \hat{E}^c\wedge \hat{E}^b \wedge \hat{E}^a\epsilon_{abcd}
 \hat{{\theta}}{\sigma}^{d} \hat{\bar{\theta}}\, \delta^4(x-\hat{x})+ {\cal O}(f^4)
 \; , \qquad
\end{eqnarray}
where $ {\cal O}(f^4)$ denotes the terms of the fourth order in fermions (in this case, these are worldvolume fermionic fields $\hat{\theta}$, $\hat{\bar{\theta}}$ and their worldvolume derivatives,  $\partial_m\hat{\theta}:=\partial \hat{\theta}/\partial\xi^m$ and {\it c.c.}); the explicit form of these one can find in the Appendix B (Eq.   (\ref{JWZ0=App})).

Substituting the above expressions  into Eq. (\ref{Eqm0=Jv0}), one finds that the real part of the auxiliary fields of the chiral multiplet has quite a complex form in terms of supermembrane variables
\begin{eqnarray}\label{Eqm0=Jv0+O}
 DD\Phi \vert_0 +\bar{D}\bar{D}\bar{\Phi} \vert_0 = {1\over 16} \,\sqrt{{\phi\over \bar{\phi}}}\; \int d^3 \xi \, \sqrt{g} \, \hat{\bar{\theta}}\hat{\bar{\theta}} \delta^4(x-\hat{x})  + {1\over 16} \sqrt{{\bar{\phi}\over \phi}}\; \int d^3 \xi \, \sqrt{g}\, \hat{{\theta}}\hat{{\theta}}\delta^4(x-\hat{x}) + \nonumber \\
 + {1\over 48}  \int_{W^3} \hat{E}^c\wedge \hat{E}^b \wedge \hat{E}^a\epsilon_{abcd}
 \hat{{\theta}}{\sigma}^{d} \hat{\bar{\theta}}\, \delta^4(x-\hat{x}) + {\cal O}(f^4)\; ,  \qquad
\end{eqnarray}
where $ {\cal O}(f^4)$ are the same as in Eq. (\ref{JWZ0=}) (and thus can be read off Eq. (\ref{JWZ0=App})).

The second auxiliary field equation, Eq. (\ref{d(aux-aux)=}), reads  \begin{eqnarray}\label{d(aux-aux)=b+}
 \partial_a( DD\Phi \vert_0 -\bar{D}\bar{D}\bar{\Phi} \vert_0 ) = - {i\over 8\cdot 4!} \int\limits_{W^3}
\hat{E}^d\wedge \hat{E}^c\wedge \hat{E}^b \epsilon_{abcd} \delta^4(x-\hat{x}) + {\cal O}(f^2) \;  , \qquad
\end{eqnarray}
where the terms of higher order in fermions, ${\cal O}(f^2)$ can be found in Eqs. (\ref{DbDJWZ0=f2}) and (\ref{DbDJWZ0=f4}) of Appendix B (multiplying the expressions presented there by $- {i\over 4} \tilde{\sigma}_a^{\dot{\alpha}\alpha}$). On the first look it might seem that Eq. (\ref{d(aux-aux)=b+}) imposes additional restrictions on the supermembrane motion. Such possible restrictions might come from the selfconsistency condition of Eq. (\ref{d(aux-aux)=b+}); at zero order in fermions that reads \footnote{
$ \int\limits_{W^3}
\hat{E}^d\wedge \hat{E}^c\wedge \hat{E}^b \epsilon_{abcd} \delta^4(x-\hat{x})= \int\limits_{W^3} d\hat{x}^d\wedge d\hat{x}^c\wedge d\hat{x}^b \epsilon_{abcd} \delta^4(x-\hat{x})+ fermionic \; contributions $.  }
\begin{eqnarray}\label{dd(aux-aux)=0}
 \partial_{[a} \epsilon_{b]c_1c_2c_3} \int\limits_{W^3} d\hat{x}^{c_3}\wedge d\hat{x}^{c_2}\wedge d\hat{x}^{c_1}  \delta^4(x-\hat{x})=0
  \;  . \qquad
\end{eqnarray}
However, one can check that  this equation is satisfied identically. Indeed, using the identity $ \epsilon_{c_1c_2c_3[a} \partial_{b]}\equiv   -{3\over 2}\epsilon_{ab[c_1c_2}\partial_{c_3]} $ one can write the {\it l.h.s.} of Eq. (\ref{dd(aux-aux)=0}) in the form of $-{3\over 2}  \epsilon_{abc_1c_2}\int\limits_{W^3} d\hat{x}^{c_2}\wedge d\hat{x}^{c_1}\wedge d \delta^4(x-\hat{x})=
-{3\over 2}  \epsilon_{abc_1c_2}\int\limits_{W^3} d\left(d\hat{x}^{c_2}\wedge d\hat{x}^{c_1}\,  \delta^4(x-\hat{x}) \right)$ which vanishes as an integral of total derivative in the case of closed supermembrane which we are studying in this paper ($\partial W^3 =0\!\!\!/\quad \Rightarrow\quad \int_{W^3}d(...)=0$) .

As we have discussed in Sec. 4.1, the on-shell transformations  are obtained from the off-shell ones,
$\delta \psi_\alpha = 2(\sigma^a\bar{\varepsilon})_\alpha \partial_a \phi + {i\over 2}\varepsilon_\beta\, DD\Phi\vert_0$ for the case of fermions, by substituting the solution of the equations for the auxiliary fields. This implies that the on-shell supersymmetry transformation of fermions will be quite complicated due to the complicated structure of the auxiliary field equations
(\ref{Eqm0=Jv0+O}) and  (\ref{d(aux-aux)=}).
As the on-shell fermionic supersymmetry transformations can be used to extract BPS conditions for the supersymmetric solutions, their further study in our simple system might lead to useful suggestions for the investigation of the backreaction of D=10,11 super-$p$-branes on the  BPS solutions of supergravity equations.

\subsubsection{Dynamical field equations}

Fortunately, the dynamical equations for the physical fields of the special scalar multiplet following form the simplest interacting action Eq. (\ref{S=S0+SM2}) do not obtain contributions from the auxiliary fields of the scalar multiplet,
which on the mass shell are expressed by quite complicated Eqs. (\ref{Eqm0=Jv0+O}) and (\ref{d(aux-aux)=}).

Specifying the current contributions to (\ref{Dirac=J}) and (\ref{K-G=J}), we find the massless Dirac equation with the supermembrane contributions,
\begin{eqnarray}\label{Dirac=J+O}
{\sigma}^a_{\alpha\dot{\alpha}}\partial_a\psi^\alpha &=&
 {1\over  32}\,\sqrt{{\phi\over \bar{\phi}}}\; \int d^3 \xi \, \sqrt{g} \, \hat{\bar{\theta}}_{\dot{\alpha}}\delta^4 (x-\hat{x})- \qquad \nonumber \\
&& - {1\over 8\cdot 4!}  \int_{W^3} \hat{E}^c\wedge \hat{E}^b \wedge \hat{E}^a\epsilon_{abcd}
 (\hat{{\theta}}{\sigma}^{d})_{\dot{\alpha}}\, \delta^4(x-\hat{x}) +{\cal O}(f^3)
   \; , \qquad
\end{eqnarray}
and the Klein-Gordon equation, also with the source from supermembrane,
\begin{eqnarray}\label{K-G=J+O}
&& \Box \phi(x)=   {1\over 64} \,\sqrt{{{\phi}\over {\bar{\phi}}}}\; \int d^3 \xi \, \sqrt{g} \delta^4 (x-\hat{x}) - \qquad \nonumber \\
&& -{1\over 64}\int\limits_{W^3} \hat{E}^c\wedge \hat{E}^b\wedge d\hat{\theta} \sigma_{bc}\hat{\theta} \, \delta^4(x-\hat{x}) -
 {i\over 64}  \int\limits_{W^3} \hat{E}^c\wedge \hat{E}^b \wedge \hat{E}^a\epsilon_{abcd} (\hat{\theta})^2\partial^d\delta^4(x-\hat{x})+ \qquad \nonumber \\ &&
+ {i\over 64} \, \int d^3 \xi \, \sqrt{g}  (\hat{{\theta}}\sigma^a\hat{\bar{\theta}}) \sqrt{{\hat{\phi}\over {\hat{\bar{\phi}}}}}\; \partial_a\delta^4 (x-\hat{x}) + {\cal O}(f^4)
\;  . \qquad
\end{eqnarray}
The explicit form of the terms of higher order in fermions, ${\cal O}(f^4)$ in  (\ref{K-G=J+O}) and
${\cal O}(f^3)$ in  (\ref{Dirac=J+O}), can be extracted from the Eqs.  (\ref{bDbDJNG0=App-2}) and (\ref{bDbDJWZ0=}) in Appendix B.

\subsection{Simplest solution of the dynamical equations at leading order in supermembrane tension}

The above equations can be formally solved by
\begin{eqnarray}\label{psi=J+O}
\psi^\alpha &=& \psi_0^\alpha
 +{1\over  32}\, \int d^3 \xi \, \sqrt{g} \, \sqrt{{\hat{\phi}\over \hat{\bar{\phi}}}}\; (\hat{\bar{\theta}}\tilde{\sigma}^a)^{\alpha}\partial_a G_0 (x-\hat{x})+\qquad \nonumber \\
&& + {1\over 8\cdot 4!}  \int_{W^3} \hat{E}^c\wedge \hat{E}^b \wedge \hat{E}^a\epsilon_{abcd}
 (\hat{{\theta}}{\sigma}^{d}\tilde{\sigma}^a)^{\alpha}\partial_a\, G_0 (x-\hat{x})  +{\cal O}(f^3)
   \; , \qquad
\\ \label{phi=J+O}
\phi(x)&=& \phi_0(x) + {1\over 64} \, \int d^3 \xi \, \sqrt{g}  \sqrt{{\hat{\phi}\over \hat{\bar{\phi}}}}\; \left(G_0 (x-\hat{x}) + i   (\hat{{\theta}}\sigma^a\hat{\bar{\theta}}) \partial_aG_0 (x-\hat{x})\right)   - \qquad \nonumber \\
&& -{1\over 64}\int\limits_{W^3} \hat{E}^c\wedge \hat{E}^b\wedge d\hat{\theta} \sigma_{bc}\hat{\theta} G_0(x-\hat{x}) -
 {i\over 64}  \int\limits_{W^3} \hat{E}^c\wedge \hat{E}^b \wedge \hat{E}^a\epsilon_{abcd} (\hat{\theta})^2\partial^d G_0(x-\hat{x})+ \nonumber \\  && + {\cal O}(f^4)
\;  , \qquad
\end{eqnarray}
where $\psi_0^\alpha$ and $\phi_0(x)$ are solutions of the free equations and $G_0 (x-\hat{x})$ is the Green function of the free $D=4$ Klein-Gordon operator $\Box :=\partial_a\partial^a$,
\begin{eqnarray}\label{BoxG0=d}
\Box \phi_0(x)=0\; , \qquad {\sigma}^a_{\alpha\dot{\alpha}}\partial_a\psi_0^\alpha =0\,  \qquad \Box G_0 (x-\hat{x})= \delta^4(x-\hat{x})\;  .  \qquad
\end{eqnarray}
Eqs. (\ref{psi=J+O}) and (\ref{phi=J+O}) give only formal solutions as far as the pull--back of the phase of
the complex scalar superfield enters their {\it r.h.s.} through $ \sqrt{{\hat{\phi}\over \hat{\bar{\phi}}}}$ multipliers in the integrants.

Assuming the solution of the homogeneous equation to be real, $\phi_0(x)=(\phi_0(x))^*$ one can solve  Eqs. (\ref{Dirac=J+O}) and (\ref{K-G=J+O}) in the first order in the supermembrane tension $T$ (this is set to unity in our equations above and below, but can be easily restored by  $\int d^3 \xi \, \sqrt{g}  \mapsto T\int d^3 \xi \, \sqrt{g}  $ and $\int\limits_{W^3}\mapsto T\int\limits_{W^3}$). This reads (setting back $T=1$)
\begin{eqnarray}\label{psi=J+O+}
\psi^\alpha &=& \psi_0^\alpha
 +{1\over  32}\, \int d^3 \xi \, \sqrt{g} \, (\hat{\bar{\theta}}\tilde{\sigma}^a)^{\alpha}\partial_a G_0 (x-\hat{x})+ \qquad \nonumber \\
&& + {1\over 8\cdot 4!}  \int_{W^3} \hat{E}^c\wedge \hat{E}^b \wedge \hat{E}^a\epsilon_{abcd}
 (\hat{{\theta}}{\sigma}^{d}\tilde{\sigma}^a)^{\alpha}\partial_a\, G_0 (x-\hat{x})  +{\cal O}(f^3)
   \; , \qquad
\\ \label{phi=J+O+}
\phi(x)&=& \phi_0(x) + {1\over 64} \, \int d^3 \xi \, \sqrt{g} \left(G_0 (x-\hat{x}) + i   (\hat{{\theta}}\sigma^a\hat{\bar{\theta}}) \partial_aG_0 (x-\hat{x})\right)   + \qquad \nonumber \\
&& -{1\over 64}\int\limits_{W^3} \hat{E}^c\wedge \hat{E}^b\wedge d\hat{\theta} \sigma_{bc}\hat{\theta} G_0(x-\hat{x}) -
 {i\over 64}  \int\limits_{W^3} \hat{E}^c\wedge \hat{E}^b \wedge \hat{E}^a\epsilon_{abcd} (\hat{\theta})^2\partial^d G_0(x-\hat{x})+ \nonumber \\  && + {\cal O}(f^4)
\;  , \qquad  \phi_0(x)= (\phi_0(x))^* \;  . \qquad
\end{eqnarray}
The contribution of higher order in string tension would include the product of distributions
(of the type $G_0(x-\hat{x}(\xi_1))\, \delta^4(x-\hat{x}(\xi_2))$) and their accounting requires a careful study of a classical counterpart of the renormalization procedure, similar to the one developed for the radiation reaction problem \cite{Lyakhovich+Sharapov+2002} and for general relativity \cite{Infeld+Plebanski}. The generalization of such a technique for the case of $p$--brane has been developed in very recent \cite{Lechner:2010dr}.

\section{Conclusion and discussion}
\label{Conclusion}
\setcounter{equation}0

In this manuscript we studied the interaction of the four dimensional supermembrane with dynamical scalar multiplet. We have derived the complete set of the superfield equations of motion for this interacting system in  the case when scalar multiplet is described by chiral superfield of special form, namely expressed through the real pre-potential superfield. Such a {\it special} scalar multiplet has a simple dual description in terms of 3-form potential.

Although the interacting systems of dynamical D=4 ${\cal N}=1$ supergravity and superstring have been studied some years ago \cite{B+I03}, the superfield equations of motion for the dynamical system including supermembrane have not been known before. Furthermore, this paper seems to be the first study  of superfield equations for interacting system involving matter (not supergravity) superfields and a supersymmetric extended object.

Furthermore, although the supermembrane in flat superspace and in the background of ${\cal N}=1$, $D=4$ supergravity is well known \cite{BST87,4Dsuperm}, to our best knowledge, the action of supermembrane in a chiral superfield background, Eq.  (\ref{SM2+=4D}), was not written before.

The characteristic coupling of the scalar field to the membrane which appears in the bosonic limit of the action (\ref{SM2+=4D}) is similar  to the coupling presented in \cite{Tomas+}, but not identical to this. The difference is that the counterpart of our arbitrary holomorphic function $K(\phi)$ of complex field $\phi$ ($\partial_{\bar{\phi}} K:= \partial K/\partial \bar{\phi}=0$) is replaced in \cite{Tomas+} by certain covariantly chiral function ${\cal L}$
which is the product of holomorphic function, superpotential $W(\phi)$, and of the exponent of the K\"{a}hler potential ${\cal K}(\phi, \bar{\phi})$ (${\cal D}_{\bar{\phi}}{\cal L}=0$ with suitable covariant derivative ${\cal D}_{\bar{\phi}}=\partial_{\bar{\phi}} +...$) . In contrast, we do not find any relation of $K(\phi)$ with superpotential $W(\Phi)$ and K\"{a}hler potential ${\cal K}$ of the most general scalar multiplet action.

This may be related to that we are studying the supermembrane in flat superspace with only the scalar supermultiplet background, requiring only the global supersymmetry and the $\kappa$--symmetry of the action, while the superfield counterpart of the membrane action  of \cite{Tomas+} should appear as a part of the interacting  superfield action including the superfield supergravity, scalar supermultiplets and supermembrane.
The construction of such an action and deriving superfield equations from it will be the subject of our future study.

A complete but gauge fixed version of such an  action, obtained on the line of \cite{BdAI1}-\cite{IB+JdA05}, should then include the pure bosonic membrane coupled to spacetime fields of the supergravity and matter multiplets, possesses invariance under the local supersymmetry 1/2 of which should be broken on the membrane worldvolume by the condition identifying the pull--back of the local supersymmetry parameter with the $\kappa$--symmetry. Just such a (hypothetical) gauge fixed action would produce the equations of \cite{Tomas+} invariant under the local supersymmetry restricted by a $\kappa$--symmetry--like projection condition on the membrane worldvolume. Such a relation of the supersymmetry with $\kappa$--symmetry,  characteristic for \cite{Tomas+} but absent in our dynamical system, might be the reason of that the counterpart of $K(\phi)$ is expressed through the superpotential and K\"{a}hler  potential in  \cite{Tomas+} while remains arbitrary holomorphic function in our case.

In Sec. 5 we have specified our bulk superfield equations for the simplest case when the bulk part of the action is given by the free kinetic term
(with K\"{a}hler potential ${\cal K}= \Phi\bar{\Phi}$ and vanishing superpotential, $W(\Phi)=0$),  extracted from these the equations for spacetime, component fields and  presented a solution of the dynamical
equations for physical fields in the leading order on supermembrane tension. A search for  solution of the superfield equations with a nontrivial superpotential is an interesting direction for future study.

\bigskip

\subsection{Discussion added}

As a first stage in this direction, let us consider the relation with known domain wall solutions of
the Wess--Zumino model \cite{Abraham+Townsend=1990,Cvetic+=1991}. To this end let us consider our dynamical system with nontrivial superpotential and the  simplest kinetic term. This is described by the interacting action
\begin{eqnarray}\label{S=SWZ+SM2}
S = \int d^8z\, \Phi \bar{\Phi}  + \int d^6\zeta_L W(\Phi) + \int d^6\zeta_R \bar{W}(\bar{\Phi})  + {1\over 2}\int d^3 \xi \sqrt{\hat{\Phi}\hat{\bar{\Phi}}}\sqrt{ g} -   \int\limits_{W^3} \hat{C}_3{}^\prime\;  .   \qquad
\end{eqnarray}
with $\Phi$ and $\bar{\Phi}$ and $\hat{C}_3{}^\prime$ expressed in terms of real pre-potential $V(z)$ by Eqs. (\ref{Phi=DDV}) and (\ref{C3'=}).

The superfield equations of motion (\ref{Eqm0=J}) acquire now the superpotential contributions,
\begin{eqnarray}\label{Eqm0+W=J}
  DD\Phi+\bar{D}\bar{D}\bar{\Phi} + W^\prime_\Phi (\Phi) + \bar{W}^\prime_{\bar{\Phi}}(\bar{\Phi}) = J(z)\;   \qquad
\end{eqnarray}
The auxiliary field equations read
\begin{eqnarray}\label{EqmW=Jv0}
&& DD\Phi \vert_0 +\bar{D}\bar{D}\bar{\Phi}\vert_0  + W^\prime_\phi (\phi) + \bar{W}^\prime_{\bar{\phi}}(\bar{\phi}) = J(z)\vert_0\; ,  \qquad \\
 \label{d(aux-aux)=W}
 && \partial_a( DD\Phi \vert_0 + \bar{W}^\prime_{\bar{\phi}}(\bar{\phi}) -\bar{D}\bar{D}\bar{\Phi} \vert_0  - W^\prime_\phi (\phi) ) = - {i\over 4} \tilde{\sigma}_a^{\dot{\alpha}\alpha}[D_\alpha \, , \, \bar{D}_{\dot\alpha}] J(z)\vert_0\;  , \qquad
\end{eqnarray}
and the dynamical field equations are
\begin{eqnarray}\label{Dirac-W=J}
&& {\sigma}^a_{\alpha\dot{\alpha}}\partial_a\psi^\alpha + {i\over 4} \bar{\psi}{}_{\dot{\alpha}} \, \bar{W}^{\prime\prime}_{\bar{\phi}\bar{\phi}}(\bar{\phi}) = {1\over 4} \bar{D}_{\dot{\alpha}}J(z)\vert_0 \; , \qquad
\\
\label{K-G-W=J}
&& \Box \phi(x) - {1\over 16} \bar{D}\bar{D}\bar{\Phi}\vert_0 \bar{W}^{\prime\prime}_{\bar{\phi}\bar{\phi}}(\bar{\phi}) +
{1\over 16} \bar{\psi}{}_{\dot{\alpha}}\bar{\psi}{}^{\dot{\alpha}} \bar{W}^{\prime\prime\prime}_{\bar{\phi}\bar{\phi}\bar{\phi}}(\bar{\phi})  = -{1\over 16} \bar{D}\bar{D}J(z)\vert_0 \;  \qquad
\end{eqnarray}
with the same supermembrane current (\ref{J=JNG+JWZ}), (\ref{JNG=}), (\ref{JWZ=}).

{\it In the absence of supermembrane current, $J=0$}, the auxiliary field equations are solved by
$DD\Phi \vert_0 =- \bar{W}^\prime_{\bar{\phi}}(\bar{\phi}) +ic$,
$\bar{D}\bar{D}\bar{\Phi}\vert_0 =- W^\prime_\phi (\phi) -ic$ and the constant $c$ can be removed by redefining superpotential $W^\prime_\phi (\phi)\mapsto W^\prime_\phi (\phi)+ ic $. Thus, without lost of generality, one can simplify notation and substitute $- \bar{W}^\prime_{\bar{\phi}}(\bar{\phi})$ for $DD\Phi \vert_0$ in the dynamical equations (\ref{K-G-W=J}).
Domain wall ansatz of \cite{Abraham+Townsend=1990,Cvetic+=1991} implies that all the fields are static and depend on only one spatial coordinate which we chose to be $x^2=y$.  Then Eqs. (\ref{Dirac-W=J}) and (\ref{K-G-W=J}) with $J=0$ becomes
\begin{eqnarray}\label{Dirac-Wy=J}
&& {\sigma}^2_{\alpha\dot{\alpha}} \partial_y \psi^\alpha (y)+ {i\over 4} \bar{\psi}{}_{\dot{\alpha}}(y) \, \bar{W}^{\prime\prime}_{\bar{\phi}\bar{\phi}}(\bar{\phi}(y)) = 0 \; , \qquad
\\
\label{K-G-Wy=J}
&& \partial_y^2 \phi(y) - {1\over 16} {W}^{\prime}_{{\phi}}({\phi})  \bar{W}^{\prime\prime}_{\bar{\phi}\bar{\phi}}(\bar{\phi})-
{1\over 16} \bar{\psi}{}_{\dot{\alpha}}\bar{\psi}{}^{\dot{\alpha}} \bar{W}^{\prime\prime\prime}_{\bar{\phi}\bar{\phi}\bar{\phi}}(\bar{\phi})  = 0 \;  \qquad
\end{eqnarray}
Notice that Eq. (\ref{Dirac-Wy=J}) split into the pair of equations for $\psi_1 ,\bar{\psi}_1$ and $\psi_2 ,\bar{\psi}_2$,
\begin{eqnarray}\label{Dirac-Wy1=J}
\partial_y \psi_1 (y)+ {1\over 4} \bar{\psi}{}_{\dot{1}}(y) \, \bar{W}^{\prime\prime}_{\bar{\phi}\bar{\phi}}(\bar{\phi}(y)) = 0 \; , \qquad \partial_y \psi_2 (y)+ {1\over 4} \bar{\psi}{}_{\dot{2}}(y) \, \bar{W}^{\prime\prime}_{\bar{\phi}\bar{\phi}}(\bar{\phi}(y)) = 0 \;
\; , \qquad
\end{eqnarray}
such that the solution of the second can be constructed from the solution of the first as $\psi_2= \psi_1$, $
\bar{\psi}{}_{\dot{2}}=\bar{\psi}{}_{\dot{1}}$.  For such a solution of the fermionic equation $\bar{\psi}{}_{\dot{\alpha}}\bar{\psi}{}^{\dot{\alpha}}=0$ and the bosonic equation simplifies to
$\partial_y^2 \phi(y) - {1\over 16} {W}^{\prime}_{{\phi}}({\phi})  \bar{W}^{\prime\prime}_{\bar{\phi}\bar{\phi}}(\bar{\phi})=0$. This, in its turn, is solved by any solution of the following first order BPS equations \cite{Abraham+Townsend=1990,Cvetic+=1991}
\begin{eqnarray}
\label{dphi=}
&& \partial_y\phi(y) - {e^{i\alpha}\over 4} \bar{W}^{\prime}_{\bar{\phi}}(\bar{\phi}(y))  = 0 \; , \qquad
\partial_y\bar{\phi}(y)- {e^{-i\alpha}\over 4} {W}^{\prime}_{{\phi}}({\phi}(y))  = 0 \; . \qquad
\end{eqnarray}
Abraham and Townsend \cite{Abraham+Townsend=1990} studied the intersecting domain wall solutions of
(\ref{dphi=}) with $W=\Phi^4-4\Phi$. The generalization of the above equations for the case of supergravity was studied in \cite{Cvetic+=1991}. For $W= a^2\Phi - {\Phi^3\over 3}$  Eq. (\ref{dphi=}) has  kink solution $\phi= a\, \tanh (ya)$ \cite{Cvetic+=1991}. Notice that in the case of special chiral multiplet such a potential would be deformed by the contribution of an arbitrary imaginary constant, $a^2 \mapsto a^2+ic$.

When the BPS equations (\ref{dphi=})  are satisfied, the solution of fermionic equations can be written in the form
\cite{Abraham+Townsend=1990}
\begin{eqnarray}
\label{psi1=}
\psi_1 =2 \chi \, e^{-i\alpha /2} \partial_y\phi(y) =\psi_2 \; ,  \qquad  \bar{\psi}_{\dot{1}} = 2 \chi e^{i\alpha /2} \partial_y\phi(y) =\bar{\psi}_{\dot{2}}\;  \qquad
 \;  \qquad
\end{eqnarray}
with a real Grassmann (fermionic) constant  $\chi$,
\begin{eqnarray}
\label{chi=chi*}
 \chi =\chi^* \; ,  \qquad \chi \chi=0 \; .
\end{eqnarray}

An interesting problem for future is to study the influence of the supermembrane source on the above discussed  nonsingular domain wall solutions. First observation is that, to maintain the general structure of the solution (\ref{psi1=}) of the fermionic equations, the source contribution in the {\it r.h.s.} of (\ref{Dirac-W=J}),
${\sigma}^2_{\alpha\dot{\alpha}}\partial_y\psi^\alpha (y)+ {i\over 4} \bar{\psi}{}_{\dot{\alpha}} (y)\, \bar{W}^{\prime\prime}_{\bar{\phi}\bar{\phi}}(\bar{\phi})(y) = {1\over 4} \bar{D}_{\dot{\alpha}}J(z)\vert_0$, should be proportional to the same real Grassmann constant  $\bar{D}_{\dot{\alpha}}J(z)\vert_0 \propto \chi$. This in its turn suggests the following ansatz for the fermionic coordinates functions
\begin{eqnarray}
\label{th=}
\hat{\theta}{}^{\alpha} (\xi) = u^{\alpha}(\xi) \chi \; , \qquad \hat{\bar{\theta}}{}^{\dot\alpha} (\xi) = \bar{u}^{\dot\alpha}(\xi) \chi
 \;   \qquad
\end{eqnarray}
with some bosonic functions $u^{\alpha}(\xi)= (\bar{u}^{\dot\alpha}(\xi))^*$. Such an ansatz results in that $\hat{\theta}{}^{\alpha} \hat{\theta}{}^{\beta} =0= \hat{\theta}{}^{\alpha} \hat{\bar{\theta}}{}^{\dot\alpha} $ and, hence, in that the pull--back of bosonic vielbein simplifies to $\hat{E}^a=d\hat{x}^a$. Furthermore, assuming that the normal to the supermembrane worldvolume cannot be orthogonal to the $y=x^2$ axis, we can chose the `static gauge' where $\hat{x}^0(\xi) =\xi^0=\tau$, $\hat{x}^1(\xi)=\xi^1$, $\hat{x}{}^3=\xi^2$ so that the only nontrivial bosonic coordinate function (supermembrane Goldstone field) is identified with $\hat{x}{}^2=\hat{y}(\xi )$. In this gauge
\begin{eqnarray}
\label{hEa=static}
\hat{E}^0=d\xi^0\; , \qquad \hat{E}^1=d\xi^1
 \; ,  \qquad  \hat{E}^2=d\hat{y}(\xi)= d\hat{y}(\xi^0,\xi^1,\xi^2)\; ,  \qquad  \hat{E}^3=d\xi^2 \; . \qquad
\end{eqnarray}
Furthermore, with such an ansatz $J\vert_0=0$ and all the (quite complicated) components of current superfield (see Appendix B) simplify drastically reducing to their leading terms. Then the problem of finding (particular) solutions of the system of interacting equations looks manageable. We hope to address this problem in the future publication.

\bigskip

\subsection*{Acknowledgments}

One of the authors (I.B.) thanks Dima Sorokin, Ulf Lindstr\"om and Martin Ro\v{c}ek  for useful conversations, Warren Siegel for useful comments and communications and the organizers of the VIII Simons Workshop,  especially Martin Ro\v{c}ek, for their kind hospitality in Stony Brook where a part of this work was done. The partial support by the research grants FIS2008-1980 from the Spanish MICINN  and by the Basque Government Research Group Grant ITT559-10 is greatly acknowledged.

\newpage

\bigskip
\section*{Appendix A: Notation,  conventions and some  useful formulae}
\renewcommand{\theequation}{A.\arabic{equation}}
\setcounter{equation}{0}

We use  mostly minus Minkowski metric $\eta^{ab}=diag (1,-1,-1,-1)$ and complex Weyl spinor notation.
$D=4$ vector and spinor indices are denoted by symbols from the beginning of Latin and Greek alphabets,
$a,b,c= 0,1,2,3$, $\alpha ,\beta , \gamma =1,2$, $\dot{\alpha}, \dot{\beta}, \dot{\gamma}=1,2$.
In particular, the coordinates of the target $D=4$, ${\cal N}=1$ superspace $\Sigma^{(4|4)}$ are denoted, respectively, by
$x^a$ and  $\theta^\alpha$,  $\bar{\theta}{}^{\dot{\alpha}} =(\theta^\alpha)^*$. The star $^*$ denotes complex conjugation of bosonic variables and involution on Grassmann algebra (see \cite{Ogievetsky:1975nu} and refs. therein); in practical terms this implies that  $(\theta^\alpha\hat{\theta}{}^\beta)^*= \hat{\bar{\theta}}{}^{\dot{\beta}} \bar{\theta}{}^{\dot{\alpha}} =-
\bar{\theta}{}^{\dot{\alpha}}\hat{\bar{\theta}}{}^{\dot{\beta}}$. Then, to keep the plus sign in $(\theta^2)^*= \bar{\theta}{}^2$ with $(\theta)^2:=\theta^\alpha\theta_\alpha$, we have to define $\bar{\theta}{}^2:= \bar{\theta}_{\dot{\alpha}} \bar{\theta}{}^{\dot{\alpha}}$. The consistency of the Grassmann algebra involution  that $(\partial_\alpha)^*\equiv  \left({\partial\over \partial {\theta}{}^{{\alpha}}}\right)^*=  -\partial_{\dot{\alpha}}\equiv - {\partial\over \partial \bar{\theta}{}^{\dot{\alpha}}}$. Then, the covariant spinor derivative defined in (\ref{Dalpha=}) are related by $(D_\alpha)^*=  -\bar{D}_{\dot{\alpha}}$. Then
$\bar{D}\bar{D}:= \bar{D}_{\dot{\alpha}}\bar{D}{}^{\dot{\alpha}}= (DD)^*$ where $DD:=D^\alpha D_\alpha$.

 The contraction the spinorial indices are raised and lowered by the unit antisymmetric tensors $\epsilon^{\alpha\beta}=-\epsilon^{\beta\alpha}=i\sigma^2\equiv \left(\matrix{0 & 1\cr -1 & 0}\right) $, $\epsilon^{\dot{\alpha}\dot{\beta}}=-\epsilon^{\dot{\beta}\dot{\alpha}}$ and their inverse
 $\epsilon_{\alpha\beta}=-\epsilon_{\beta\alpha}$, $\epsilon_{\dot{\alpha}\dot{\beta}}=-\epsilon_{\dot{\beta}\dot{\alpha}}$. This implies $\theta_\alpha= \epsilon_{\alpha\beta}\theta^\beta$, $\theta^\alpha= \epsilon^{\alpha\beta}\theta_\beta$, etc. However, to get $\partial^\alpha \theta_\beta =\delta_\beta{}^\alpha$ simultaneously with  $\partial_\alpha \theta^\beta =\delta_\alpha{}^\beta$ we have to assume that for the derivatives over the fermionic variables
 $\partial^\alpha = - \epsilon^{\alpha\beta}\partial_{\beta}$ so that, when we rise the spinorial index of the covariant fermionic derivative (\ref{Dalpha=}), we arrive at  $D^\alpha:=  \epsilon^{\alpha\beta}D_{\beta}=
 - \partial^\alpha - i (\bar{\theta}\tilde{\sigma}{}^a)^\alpha \partial_a$.

The list of properties of relativistic Pauli matrices $\sigma^a_{\beta\dot{\alpha}}= \epsilon_{\beta\alpha} \epsilon_{\dot{\alpha}\dot{\beta}}
 \tilde{\sigma}{}^{a\dot{\beta}\alpha}$ include
\begin{eqnarray}\label{sasb=}
& \sigma^a\tilde{\sigma}{}^b =\eta^{ab} +{i\over 2}\epsilon^{abcd}\sigma_c\tilde{\sigma}_d\; ,\qquad \tilde{\sigma}{}^a{\sigma}^b =\eta^{ab} -{i\over 2}\epsilon^{abcd}\tilde{\sigma}_c{\sigma}_d\; ,\qquad
\\ \label{sab=}
& \sigma^{ab}:={1\over 2}(\sigma^a\tilde{\sigma}{}^b-\sigma^b\tilde{\sigma}{}^a)= {i\over 2}\epsilon^{abcd}\sigma_{cd}\; , \qquad \\ \label{tsab=} & \tilde{\sigma}{}^{ab} :={1\over 2}(
\tilde{\sigma}{}^a{\sigma}^b-\tilde{\sigma}{}^b{\sigma}^a)= -{i\over 2}\epsilon^{abcd}\tilde{\sigma}_{cd}\; ,  \qquad \\
\label{sss=}
& \sigma^{abc}= -i \epsilon^{abcd}{\sigma}_d\; .  \qquad
\end{eqnarray}

\bigskip

The 3-dimensional worldvolume vector indices are denoted by symbols from the middle of Latin alphabet. In particular, the local coordinates of the supermembrane worldvolume $W^3$ are denoted by $\xi^m$ with $m=0,1,2$.
The worldvolume Hodge star operation is defined as  in (\ref{*Ea:=}),
\begin{eqnarray}\label{*Ea:=APP}
*\hat{E}^a:= {1\over 2}d\xi^m\wedge d\xi^n\sqrt{g}\epsilon_{mnk}g^{kl}\hat{E}_l^a \; . \qquad
\end{eqnarray}
In our conventions $d\xi^m\wedge d\xi^n\wedge d\xi^k= - \epsilon^{mnk} d^3\xi \equiv  \epsilon^{knm} d^3\xi$
so that
\begin{eqnarray}\label{*EaEa:=}
*\hat{E}_a\wedge \hat{E}^a =- 3 d^3\xi \sqrt{g} \; , \qquad
*\hat{E}_a\wedge \delta \hat{E}^a =-  d^3\xi \sqrt{g} \hat{E}_{ma} g^{mn}\delta \hat{E}_n^a\;  \qquad
\end{eqnarray}
and
\begin{eqnarray}\label{v*EaEa:=}
\delta(*\hat{E}_a\wedge \hat{E}^a) =3
*\hat{E}_a\wedge \delta \hat{E}^a \; . \qquad
\end{eqnarray}
The superspace generalization of Dirac delta function reads $$\delta^8 (z):= {1\over 16}(\theta)^2(\bar{\theta})^2 \delta^4 (x)$$
and obeys
$$
\int d^8 z\delta^8 (z-\hat{z}) f(z) = f(\hat{z}) \; , \qquad \int d^8 z\delta^8 (z)= \int d^4x \bar{D}\bar{D}DD \delta^8 (z)=1 $$

\section*{Appendix B:  Components of the supermembrane current superfield $J(z)=J^{NG}(z)+J^{WZ}(z)$}
\renewcommand{\theequation}{B.\arabic{equation}}
\setcounter{equation}{0}

The leading component of the Nambu-Goto current (\ref{JNG=}) reads
\begin{eqnarray}\label{JNG0=App}
 J^{NG}\vert_{0} &=& + {1\over 16} \,\sqrt{{\phi\over \bar{\phi}}}\; \int d^3 \xi \, \sqrt{g} \, \hat{\bar{\theta}}\hat{\bar{\theta}} \delta^4(x-\hat{x})  + {1\over 16} \sqrt{{\bar{\phi}\over \phi}}\; \int d^3 \xi \, \sqrt{g}\, \hat{{\theta}}\hat{{\theta}}\delta^4(x-\hat{x})   \;  \qquad \end{eqnarray}
The general expression for the fermionic derivative of this current superfield is
\begin{eqnarray} \label{bDJNG=App}
 \bar{D}_{\dot{\alpha}} J^{NG} &=& - {1\over 4} \, \int d^3 \xi \, \sqrt{g}   \,\sqrt{{\hat{\Phi}\over \hat{\bar{\Phi}}}}\; \bar{D}_{\dot{\alpha}}DD \delta^8 (Z-\hat{Z})  \; . \qquad
 \end{eqnarray}
 so that
 \begin{eqnarray} \label{bDJNG0=App}
&\bar{D}_{\dot{\alpha}} J^{NG}\vert_{0} = {1\over 8}\,\sqrt{{\phi\over \bar{\phi}}}\; \int d^3 \xi \, \sqrt{g} \, \hat{\bar{\theta}}_{\dot{\alpha}}\delta^4 (x-\hat{x}) -   {i\over 16} \, \int d^3 \xi \, \sqrt{g}\,\sqrt{{\hat{\phi}\over \hat{\bar{\phi}}}}\; (\sigma^a\hat{{\theta}})_{\dot{\alpha}} (\hat{\bar{\theta}})^2\, \partial_a\delta^4 (x-\hat{x})\,
 \; . \qquad
 \end{eqnarray}

 Furthermore, as
\begin{eqnarray} \label{bDbDJNG=App}
 \bar{D}\bar{D} J^{NG} &=& - {1\over 4}  \int d^3 \xi \, \sqrt{g} \, \,\sqrt{{\hat{\Phi}\over \hat{\bar{\Phi}}}}\; \bar{D}\bar{D} DD \delta^8 (Z-\hat{Z})    \qquad
 \end{eqnarray}
we find that
\begin{eqnarray} \label{bDbDJNG0=App}
&& \bar{D}\bar{D} J^{NG}\vert_0 = \qquad \nonumber \\ &&= \!\!  \int d^3 \xi \, \sqrt{g} \,\sqrt{{\hat{\phi}\over \hat{\bar{\phi}}}}\; \left(-{1\over 4}\delta^4 (x-\hat{x}) -{i\over 4} (\hat{{\theta}}\sigma^a\hat{\bar{\theta}}) \partial_a\delta^4 (x-\hat{x})+{1\over 16} (\hat{{\theta}})^2(\hat{\bar{\theta}})^2 \Box\delta^4 (x-\hat{x}) \right) = \qquad \nonumber \\
&&= -{1\over 4} \,\sqrt{{{\phi}\over {\bar{\phi}}}}\; \int d^3 \xi \, \sqrt{g} \delta^4 (x-\hat{x})
-  \qquad \nonumber \\ &&
-{i\over 4} \,\sqrt{{{\phi}\over {\bar{\phi}}}}\; \int d^3 \xi \, \sqrt{g}  (\hat{{\theta}}\sigma^a\hat{\bar{\theta}}) \left(\partial_a\delta^4 (x-\hat{x})+ \left({\partial_a\phi \over 2\phi }  - {\partial_a\bar{\phi}\over 2 \bar{\phi}}  \right)\delta^4 (x-\hat{x})   \right) +{\cal O}(f^4)\; . \qquad
  \end{eqnarray}
One can also write  Eq. (\ref{bDbDJNG0=App}) in the equivalent but more compact form of
\begin{eqnarray} \label{bDbDJNG0=App-2} \bar{D}\bar{D} J^{NG}\vert_0&=& -{1\over 4}\,\sqrt{{\phi\over \bar{{\phi}}}}\; \int d^3 \xi \, \sqrt{g} \, \delta^4 (x-\hat{x})  + {\cal O}(f^2)_{[\bar{D}\bar{D} J^{NG}]} + {\cal O}(f^4)_{[\bar{D}\bar{D} J^{NG}]}  \; ,  \qquad
 \end{eqnarray}
  where
 \begin{eqnarray} \label{bDbDJNG0=App(f2)}
& {\cal O}(f^2)_{[\bar{D}\bar{D} J^{NG}]} = -{i\over 4} \,\sqrt{{{\phi}\over {\bar{\phi}}}}\; \int d^3 \xi \, \sqrt{g}  (\hat{{\theta}}\sigma^a\hat{\bar{\theta}}) \left(\partial_a\delta^4 (x-\hat{x})+ \left({\partial_a\phi \over 2\phi }  - {\partial_a\bar{\phi}\over 2 \bar{\phi}}  \right)\delta^4 (x-\hat{x})   \right)  \; . \qquad
 \end{eqnarray}

 The leading term of the Wess--Zumino (WZ) contribution to the supermembrane current reads
\begin{eqnarray}\label{JWZ0=App}
  J^{WZ}(x) := J^{WZ}(Z) \vert_0&=&  {1\over 48}  \int_{W^3} \hat{E}^c\wedge \hat{E}^b \wedge \hat{E}^a\epsilon_{abcd}
 \hat{{\theta}}{\sigma}^{d} \hat{\bar{\theta}}\, \delta^4(x-\hat{x})
 - \nonumber \\ &&  - {1\over 16} \int_{W^3} \hat{E}^c\wedge \hat{E}^b\wedge d\hat{{\theta}}{}^\alpha \sigma_{bc\alpha}{}^{\beta}\hat{{\theta}}_\beta \, \hat{\bar{\theta}}\hat{\bar{\theta}} \, \delta^4(x-\hat{x})  -  \nonumber \\ &&  -  {1\over 16} \int_{W^3} \hat{E}^c\wedge \hat{E}^b \wedge d\hat{\bar{\theta}}{}^{\dot\alpha} \tilde{\sigma}_{bc}{}^{\dot{\beta}}{}_{\dot\alpha} \hat{\bar{\theta}}_{\dot{\beta}} \hat{{\theta}}\hat{{\theta}}\,  \delta^4(x-\hat{x}) +  \nonumber \\ && + {i\over 8}\int_{W^3} \hat{E}^c\wedge d\hat{{\theta}}{}^\alpha \wedge d\hat{\bar{\theta}}{}^{\dot\alpha} \sigma_{c\alpha\dot\alpha}  \hat{{\theta}}\hat{{\theta}}\, \hat{\bar{\theta}}\hat{\bar{\theta}} \, \delta^4(x-\hat{x}) = \qquad\nonumber \\ &=&  {1\over 48}  \int_{W^3} \hat{E}^c\wedge \hat{E}^b \wedge \hat{E}^a\epsilon_{abcd}
 \hat{{\theta}}{\sigma}^{d} \hat{\bar{\theta}}\, \delta^4(x-\hat{x}) +{\cal O}(f^4)
 \; . \qquad
\end{eqnarray}

Its closest fermionic partners read
\begin{eqnarray}\label{bDJWZ0=App}
 {D}_{{\alpha}}J^{WZ} \vert_0 &=&  -{1\over 48}  \int_{W^3} \hat{E}^c\wedge \hat{E}^b \wedge \hat{E}^a\epsilon_{abcd}
 ({\sigma}^{d}\hat{\bar{\theta}})_{{\alpha}}\, \delta^4(x-\hat{x})
  +\nonumber \\
  &&   + {1\over 16}  \int_{W^3} \hat{E}^c\wedge \hat{E}^b\wedge \left( 2(d\hat{\bar{\theta}} \tilde{\sigma}_{bc}\hat{\bar{\theta}}) \, \hat{{\theta}}_{{\alpha}} \,+   d\hat{{\theta}}{}^\beta {\sigma}_{bc}{}_{\beta {\alpha}} (\hat{{\theta}})^2\, \right) \delta^4(x-\hat{x}) + \nonumber \\
  && + {i\over 96}  \int_{W^3} \hat{E}^c\wedge \hat{E}^b \wedge \hat{E}^a\epsilon_{abcd}
 ({\sigma}^{de}\hat{{\theta}})_{{\alpha}}  (\hat{\bar{\theta}})^2 \partial_e  \delta^4(x-\hat{x})+ \nonumber \\ &&   + {i\over 32}\int_{W^3} \hat{E}^c\wedge \hat{E}^b\wedge d{\hat{\bar{\theta}}}{}^{\dot{\alpha}}(\sigma^a\tilde{\sigma}_{bc})_{\alpha\dot{\alpha}} (\hat{{\theta}})^2(\hat{\bar{\theta}})^2
 \partial_a \delta^4(x-\hat{x}) -  \nonumber \\ && - {i\over 4}\int_{W^3} \hat{E}^c\wedge d\hat{{\theta}}{}^\beta \wedge d\hat{\bar{\theta}}{}^{\dot\beta}\, \sigma_{c\beta\dot\beta}  \hat{{\theta}}_{\alpha} \,  (\hat{\bar{\theta}})^2 \,  \, \delta^4(x-\hat{x})  =  \qquad
 \nonumber \\ &=&  -{1\over 48}  \int_{W^3} \hat{E}^c\wedge \hat{E}^b \wedge \hat{E}^a\epsilon_{abcd}
 ({\sigma}^{d}\hat{\bar{\theta}})_{{\alpha}}\, \delta^4(x-\hat{x}) +{\cal O}(f^3)
 \; . \qquad
\end{eqnarray}
\begin{eqnarray}\label{bDJWZ0=App}
 \bar{D}_{\dot{\alpha}}J^{WZ} \vert_0 &=&  {1\over 48}  \int_{W^3} \hat{E}^c\wedge \hat{E}^b \wedge \hat{E}^a\epsilon_{abcd}
 (\hat{{\theta}}{\sigma}^{d})_{\dot{\alpha}}\, \delta^4(x-\hat{x})
  -\nonumber \\
  &&   - {1\over 16}  \int_{W^3} \hat{E}^c\wedge \hat{E}^b\wedge \left( 2(d\hat{{\theta}} \sigma_{bc}\hat{{\theta}}) \, \hat{\bar{\theta}}_{\dot{\alpha}} \,+   (d\hat{\bar{\theta}}\tilde{\sigma}_{bc})_{\dot{\alpha}} (\hat{{\theta}})^2\, \right) \delta^4(x-\hat{x}) - \nonumber \\
  && - {i\over 96}  \int_{W^3} \hat{E}^c\wedge \hat{E}^b \wedge \hat{E}^a\epsilon_{abcd}
 (\hat{\bar{\theta}}\tilde{\sigma}^{de})_{\dot{\alpha}}  (\hat{{\theta}})^2 \partial_e  \delta^4(x-\hat{x})- \nonumber \\ &&   - {i\over 32}\int_{W^3} \hat{E}^c\wedge \hat{E}^b\wedge (d\hat{{\theta}}\sigma_{bc}\sigma^a)_{\dot{\alpha}} (\hat{{\theta}})^2(\hat{\bar{\theta}})^2
 \partial_a \delta^4(x-\hat{x}) +  \nonumber \\ && + {i\over 4}\int_{W^3} \hat{E}^c\wedge d\hat{{\theta}}{}^\beta \wedge d\hat{\bar{\theta}}{}^{\dot\beta}\, \sigma_{c\beta\dot\beta}  \hat{\bar{\theta}}_{\dot\alpha} \,  (\hat{{\theta}})^2 \,  \, \delta^4(x-\hat{x})  =  \qquad
 \nonumber \\ &=&  {1\over 48}  \int_{W^3} \hat{E}^c\wedge \hat{E}^b \wedge \hat{E}^a\epsilon_{abcd}
 (\hat{{\theta}}{\sigma}^{d})_{\dot{\alpha}}\, \delta^4(x-\hat{x}) +{\cal O}(f^3)
 \; . \qquad
\end{eqnarray}

Then, as far as
\begin{eqnarray}\label{DDJWZ=}
{D}{D}J^{WZ}(Z) &=& 2i \int_{W^3} \hat{E}^c\wedge \hat{E}{}^\alpha \wedge \hat{E}{}^{\dot\alpha} \sigma_{c\alpha\dot\alpha} {D}{D} \delta^8(Z-\hat{Z})  +  \nonumber \\
&& - {1\over 2}  \int_{W^3} \hat{E}^c\wedge \hat{E}^b\wedge \hat{\bar{E}}{}^{\dot\alpha} \tilde{\sigma}_{bc}{}^{\dot\beta}{}_{\dot\alpha}
\left(\bar{D}_{\dot\beta}{D}{D}+ 4i \sigma^a_{\beta\dot\beta} \partial_a {D}^{{\beta}}\right) \delta^8(Z-\hat{Z}) +  \nonumber \\ &&  + {i\over 6}  \int_{W^3} \hat{E}^c\wedge \hat{E}^b \wedge \hat{E}^a\epsilon_{abcd} \partial^{d}{D}{D} \delta^8(Z-\hat{Z}) \; , \qquad
\\ \label{bDbDJWZ=}
\bar{D}\bar{D}J^{WZ}(Z) &=& 2i \int_{W^3} \hat{E}^c\wedge \hat{E}{}^\alpha \wedge \hat{E}{}^{\dot\alpha} \sigma_{c\alpha\dot\alpha} \bar{D}\bar{D} \delta^8(Z-\hat{Z})  +  \nonumber \\
&& + {1\over 2}  \int_{W^3} \hat{E}^c\wedge \hat{E}^b\wedge \hat{E}{}^\alpha \sigma_{bc\alpha}{}^{\beta}
\left(D_\beta \bar{D}\bar{D}- 4i \sigma^a_{\beta\dot\beta} \partial_a \bar{D}^{\dot{\beta}}\right) \delta^8(Z-\hat{Z}) -  \nonumber \\ &&  - {i\over 6}  \int_{W^3} \hat{E}^c\wedge \hat{E}^b \wedge \hat{E}^a\epsilon_{abcd} \partial^{d}\bar{D}\bar{D} \delta^8(Z-\hat{Z}) \; , \qquad
\end{eqnarray}
one finds that
\begin{eqnarray}\label{DDJWZ0=}
{D}{D}J^{WZ}\vert_0 &=& {1\over 4}\int\limits_{W^3} \hat{E}^c\wedge \hat{E}^b\wedge d\hat{\bar{\theta}} \tilde{\sigma}_{bc}\hat{\bar{\theta}} \delta^4(x-\hat{x}) -
 {i\over 4!}  \int\limits_{W^3} \hat{E}^c\wedge \hat{E}^b \wedge \hat{E}^a\epsilon_{abcd} (\hat{\bar{\theta}})^2\partial^d\delta^4(x-\hat{x}) -
\nonumber \\ && -{i\over 2} \int_{W^3} \hat{E}^c\wedge d\hat{\theta}{}^\alpha \wedge d\hat{\bar{\theta}}{}^{\dot\alpha} \sigma_{c\alpha\dot\alpha} (\hat{\bar\theta}){}^2 \delta^4(x-\hat{x})
  +  \nonumber \\
&& +{i\over 8} \int_{W^3} \hat{E}^c\wedge \hat{E}^b\wedge d\hat{\bar{\theta}}{}^{\dot{\alpha}} (\hat{{\theta}}\sigma^a\tilde{\sigma_{bc}})_{\dot{\alpha}}\,  (\hat{\bar\theta}){}^2 \partial_a\delta^4(x-\hat{x}) =  \nonumber \\  &=& {1\over 4}\int\limits_{W^3} \hat{E}^c\wedge \hat{E}^b\wedge d\hat{\bar{\theta}} \tilde{\sigma}_{bc}\hat{\bar{\theta}} \delta^4(x-\hat{x}) -
 \nonumber \\ && -{i\over 4}  \int\limits_{W^3} \hat{E}^c\wedge \hat{E}^b \wedge \hat{E}^a\epsilon_{abcd} (\hat{\bar{\theta}})^2\partial^d\delta^4(x-\hat{x})
  +  {\cal O}(f^4)\; . \qquad
\end{eqnarray}
\begin{eqnarray}\label{bDbDJWZ0=}
\bar{D}\bar{D}J^{WZ}\vert_0 &=& {1\over 4}\int\limits_{W^3} \hat{E}^c\wedge \hat{E}^b\wedge d\hat{\theta} \sigma_{bc}\hat{\theta} \delta^4(x-\hat{x}) +
 {i\over 4!}  \int\limits_{W^3} \hat{E}^c\wedge \hat{E}^b \wedge \hat{E}^a\epsilon_{abcd} (\hat{\theta})^2\partial^d\delta^4(x-\hat{x}) -
\nonumber \\ && -{i\over 2} \int_{W^3} \hat{E}^c\wedge d\hat{\theta}{}^\alpha \wedge d\hat{\bar{\theta}}{}^{\dot\alpha} \sigma_{c\alpha\dot\alpha} (\hat{\theta}){}^2 \delta^4(x-\hat{x})
  -  \nonumber \\
&& -{i\over 8} \int_{W^3} \hat{E}^c\wedge \hat{E}^b\wedge d\hat{\theta} \sigma_{bc}\sigma^a\hat{\bar{\theta}}\,  (\hat{\theta}){}^2 \partial_a\delta^4(x-\hat{x}) =  \nonumber \\  &=& {1\over 4}\int\limits_{W^3} \hat{E}^c\wedge \hat{E}^b\wedge d\hat{\theta} \sigma_{bc}\hat{\theta} \delta^4(x-\hat{x}) +
\nonumber \\ && +
 {i\over 4}  \int\limits_{W^3} \hat{E}^c\wedge \hat{E}^b \wedge \hat{E}^a\epsilon_{abcd} (\hat{\theta})^2\partial^d\delta^4(x-\hat{x})+ {\cal O}(f^4)\; . \qquad
\end{eqnarray}
To analyze the structure of the auxiliary field equation one needs also to know
\begin{eqnarray}\label{DbDJWZ0=}
[{D}_\alpha\, , \, \bar{D}_{\dot\beta}]J^{WZ}\vert_0 &=& -{1\over 4!}\int\limits_{W^3}
\hat{E}^c\wedge \hat{E}^b\wedge \hat{E}^a \epsilon_{abcd}\sigma^d_{\alpha\dot{\beta}} \delta^4(x-\hat{x}) + {\cal O}(f^2) \; , \qquad
\end{eqnarray}
where
\begin{eqnarray}
\label{DbDJWZ0=f2}
{\cal O}(f^2) &=& {i\over 4!}\int\limits_{W^3}
\hat{E}^c\wedge \hat{E}^b\wedge \hat{E}^a \epsilon_{abcd}\left( (\sigma^{de}\hat{\theta})_{\alpha}\hat{\bar{\theta}}_{\dot{\beta}} + \hat{{\theta}}_{\alpha}(\hat{\bar{\theta}}\tilde{\sigma}{}^{de})_{\dot{\beta}}\right) \partial_e\delta^4(x-\hat{x}) +
\nonumber \qquad \\ &&
 + {1\over 4}  \int_{W^3} \hat{E}^c\wedge \hat{E}^b\wedge \left(( \sigma_{bc}d\hat{\theta})_\alpha \hat{\bar{\theta}}_{\dot\beta}\,  - (d\hat{\bar{\theta}} \tilde{\sigma}_{bc})_{\dot\beta} \hat{{\theta}}_\alpha\,  \right)\delta^4(x-\hat{x}) + {\cal O}(f^4)\; , \qquad \\
\label{DbDJWZ0=f4}
{\cal O}(f^4) &=& {1\over 2\cdot 4!}\int\limits_{W^3}
\hat{E}^c\wedge \hat{E}^b\wedge \hat{E}^a (\hat{\theta})^2(\hat{\bar{\theta}})^2 \epsilon_{abcd}\left( \sigma^d_{\alpha\dot{\beta}}
\Box
\delta^4(x-\hat{x})  - \sigma^e_{\alpha\dot{\beta}} \partial_e \partial^d \delta^4(x-\hat{x})\right) + \nonumber \\
&& +{i\over 8} \int_{W^3} \hat{E}^c\wedge \hat{E}^b\wedge (d\hat{{\theta} }{\sigma}_{bc}{\sigma}^a)_{\dot\beta} \hat{{\theta}}_{\alpha} \,  (\hat{\bar{\theta}}){}^2 \partial_a\delta^4(x-\hat{x})  + \nonumber \\
&& +{i\over 8} \int_{W^3} \hat{E}^c\wedge \hat{E}^b\wedge (d\hat{\bar{\theta} } \tilde{\sigma}_{bc}\tilde{\sigma}^a)_{\alpha} \hat{\bar{\theta}}_{\dot{\beta}} \,  (\hat{\theta}){}^2 \partial_a\delta^4(x-\hat{x}) -   \nonumber \\ &&
 -{i} \int_{W^3} \hat{E}^c\wedge d\hat{\theta}{}^\gamma \wedge d\hat{\bar{\theta}}{}^{\dot\gamma} \sigma_{c\gamma\dot\gamma} \hat{\theta}_{\alpha} \hat{\bar{\theta}}_{\dot{\beta}} \delta^4(x-\hat{x})
 \; . \qquad
\end{eqnarray}

\end{document}